\documentclass[useAMS,usenatbib,one column]{mn2e}

\usepackage[usenames,dvipsnames]{xcolor}

\usepackage{amssymb, amsmath}
\usepackage{epsf}
\usepackage{bm}
\usepackage{ulem}

\usepackage{natbib}
\usepackage{graphicx}
\usepackage{cancel}

\def\la{\; \raise0.3ex\hbox{$<$\kern-0.75em\raise-1.1ex\hbox{$\sim$}}\;}
\def\ga{\;  \raise0.3ex\hbox{$>$\kern-0.75em\raise-1.1ex\hbox{$\sim$}}\;}

\newcommand{\aap}{Astron.\ Astrophys.}
\newcommand{\mnras}{Mon.\ Not.\ R.\ Astron.\ Soc.}
\newcommand{\physrep}{Phys.\ Rep.}
\newcommand{\apjl}{Astrophys.\ J.\ Lett.}

\newcommand{\apj}{Astrophys.\ J.}
\newcommand{\araa}{Ann.\ Rev.\ Astron.\ Astrophys.}
\newcommand{\prc}{Phys.\ Rev.\ C}
\newcommand{\prd}{Phys.\ Rev.\ D}
\newcommand{\nat}{Nature}

\newcommand{\aj}{AJ}

\def\la{\; \raise0.3ex\hbox{$<$\kern-0.75em\raise-1.1ex\hbox{$\sim$}}\;}
\def\ga{\;  \raise0.3ex\hbox{$>$\kern-0.75em\raise-1.1ex\hbox{$\sim$}}\;}

\title[Neutron stars climbing stability peaks]
{Observational signatures of neutron stars
in low-mass X-ray binaries climbing a stability peak}

\author[E. M. Kantor, M. E. Gusakov, A. I. Chugunov]
{E.~M.~Kantor$^{1}$  \thanks{kantor@mail.ioffe.ru},
M. E. Gusakov$^{1,2}$, 
A. I. Chugunov$^{1}$
\\
$^1$Ioffe Physical-Technical Institute of the Russian Academy of
Sciences,
Polytekhnicheskaya 26, 194021 St.-Petersburg, Russia
\\
$^2$Peter the Great St.Petersburg Polytechnic University,
Polytekhnicheskaya 29, 195251 St.-Petersburg, Russia
}

\begin{document}

\date{Accepted 2015 xxxx. Received 2015 xxxx;
in original form 2015 xxxx}

\pagerange{\pageref{firstpage}--\pageref{lastpage}} \pubyear{2015}

\maketitle

\label{firstpage}

\begin{abstract}
In the recent papers by \cite{gck14_short,gck14_large}
a new scenario describing evolution of rapidly rotating neutron stars
in low-mass X-ray binaries was proposed.
The scenario accounts for a resonant interaction
of normal r modes with superfluid inertial modes
at some specific internal stellar temperatures
(``resonance temperatures'').
This interaction results in an enhanced damping of
r mode and appearance of the ``stability peaks''
in the temperature -- spin frequency plane,
which split the r-mode instability window
in the vicinity of the resonance temperatures.
The scenario suggests that the hot and rapidly rotating NSs
spend most of their life climbing up these peaks and,
in particular,
are observed there at the moment.
We analyze in detail possible observational signatures of this suggestion.
In particular, we show that these objects may exhibit
`anti-glitches' -- sudden frequency jumps on a time scale of hours-months.

\end{abstract}


\maketitle

\section{Introduction}
\label{Sec:Intro}

\begin{figure}
    \begin{center}
        \leavevmode
        \includegraphics[width=6in]{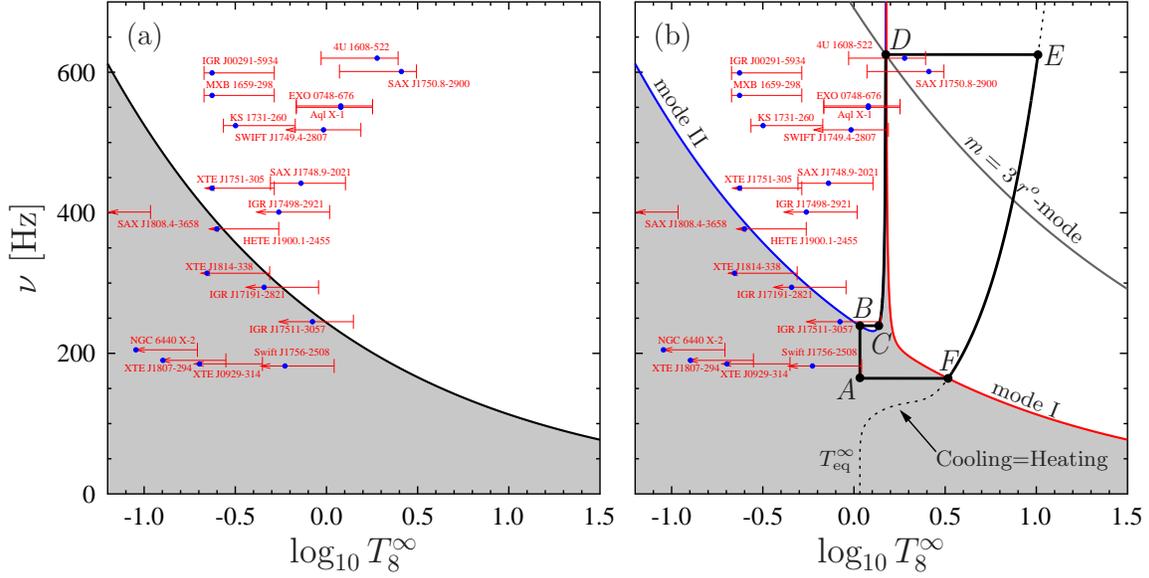}
    \end{center}
    \caption{
Panel (a) shows `standard' NS instability window (white region) in the $T^\infty-\nu$ plane,
where $T^\infty_8\equiv T^\infty/(10^8\,\rm K)$.
A region where NSs are stable (i.e.\ r-modes are not excited) is filled with grey.
Circles with error bars show observational data for NSs in LMXBs with measured $\nu$ and
estimated $T^\infty$.
Panel (b) presents
an example of the instability window (figure 5 of \citealt{gck14_large})
plotted allowing for the
resonance interaction of r mode with superfluid inertial modes.
The stability peak CD appears because of this interaction.
Evolution of the spin frequency $\nu$ and internal redshifted temperature
$T^\infty$ is shown by thick solid line (the track $A-B-C-D-E-F-A$).
See  \citealt{gck14_large} for more details.
    }
    \label{Fig:inst1}
\end{figure}

Neutron stars (NSs) are compact rotating relativistic objects.
Rotation allows NSs to harbor inertial oscillation modes,
the most interesting representatives of which are r modes.
As it was shown by \cite*{andersson98} and \cite*{fm98},
in the absence of dissipation r modes are subject
to a gravitational driven instability (the CFS instability)
at {\it any} NS spin frequency $\nu$.
An account for dissipative processes stabilizes
NSs to some extent,
resulting in the appearance of the so called ``instability window'' in the
$T^\infty$ -- $\nu$ plane, where $T^\infty$ is the redshifted internal stellar temperature.
A typical instability window is shown in panel (a) of Fig.\ \ref{Fig:inst1}.
In the region filled with gray NSs are stable (we call it ``stability region''),
in the white region they are unstable (it is ``instability window'').

Observations of NSs in low mass X-ray binaries (LMXBs)
revealed that many
NSs fall well outside the stability region
(see circles with error bars in panel (a) of Fig.\ \ref{Fig:inst1}
and \citealt*{hah11,hdh12,gck14_large,gck14_short}).
At the same time, NS evolution models predict that a probability to find
an NS in the instability window is negligibly small (\citealt*{levin99, heyl02}).
This apparent contradiction
was addressed in a number of papers (e.g., \citealt*{ak01,hdh12,hah11,ms13,hga14,ahs15})
which attempted to reconcile theory with observations
(see \citealt{hah11,hdh12,gck14_large,haskell15} for a short review of the existing ideas).
In this paper we explore observational consequences
of one such idea (\citealt{gck14_short, gck14_large}) which,
as we believe, allows one
to explain the existing observations rather naturally.

\cite{gck14_short, gck14_large}
argued that
a simultaneous  account for superfluidity of nucleons
in the cores of NSs and finite-temperature effects 
substantially modifies the NS oscillation spectra
and leads to a
resonance interaction of r mode with superfluid inertial modes
at some certain NS temperatures $T_{\rm res}^\infty$
(called ``resonance temperatures'' in what follows).
This resonance interaction stabilizes r modes in the vicinity of $T_{\rm res}^\infty$.
As a result, stability peaks appear in the $T^\infty-\nu$ plane.
A typical instability window constructed
allowing for the resonance interaction of the modes
is demonstrated in panel (b) of Fig.\ \ref{Fig:inst1}
(the figure is taken from \citealt{gck14_large} where it is discussed in detail).
The evolution track of an NS
in such instability window was studied by \cite{gck14_short,gck14_large} and is shown here by the solid thick line.
It was found by \cite{gck14_short, gck14_large}
that an NS spends substantial amount of time climbing up
the left edge of the stability peak so that the probability to find it there is high.
This presents a natural explanation for the existence of numerous NSs
in the region that was thought to be unstable with respect to r-mode excitation
--- they climb up the stability peak.

Detailed analyzes of the evolution track revealed (\citealt{gck14_large})
that it undergoes small oscillations (in what follows we will call them $\alpha$-oscillations)
near the edge of the peak (unresolved on the scale of Fig.\ \ref{Fig:inst1})
\footnote{Similar oscillations (in different context)
were discussed by \cite{levin99} and \cite*{whl01}.}.
These oscillations
can, in principle, be detected and this paper discusses their possible
observational manifestations.

The paper is organized as follows.
In Section \ref{Sec:GeneralEquations}
we present basic equations governing evolution of NSs in LMXBs,
then in Section \ref{Sec:Equations}
we analyze $\alpha$-oscillation properties of a star climbing up/down the stability peak.
In Section \ref{Sec:parameters} we estimate different parameters of $\alpha$-oscillations
(e.g., oscillation period).
Section \ref{subSec:timing} analyzes peculiar
timing behavior of an NS attached to the stability peak.
In Section \ref{subSec:gravrad} we discuss whether
gravitational radiation from such stars can be detected.
Section \ref{Sec:evidences} inspects existing observations.
We conclude in Section~\ref{Sec:Conclusions}.

\section{General equations}
\label{Sec:GeneralEquations}

The equations describing evolution of NSs in LMXBs consist of
(for more details see, e.g., \citealt{olcsva98,hl00,gck14_large}):

(i) Thermal balance equation
\begin{equation}
C_{\rm tot} \frac{dT^\infty}{dt} = W_{\rm Diss}-L_{\rm cool} + K_{\rm n}  \dot{M} c^2,
\label{thermal}
\end{equation}
which describes evolution of
internal redshifted temperature $T^\infty$ of the star
(we assume it is uniform throughout the core).
In equation (\ref{thermal})
$C_{\rm tot}$ is the total heat capacity of an NS;
$L_{\rm cool}$ is the thermal luminosity of the star
due to neutrino emission from
the
interior and photon emission from
its
surface.
The term $K_{\rm n}  \dot{M} c^2$
describes deep crustal heating due to
nuclear transformations of accreted matter (see \citealt*{bbr98});
$K_{\rm n}$ characterizes the efficiency of this heating
(following \citealt*{brown00} and \citealt{btw07} we adopt here $K_{\rm n}=10^{-3}$);
$\dot{M}$
is the accretion rate from the low-mass companion,
in what follows we will assume that it is constant and equals
to the accretion rate averaged over quiescent and active phases, $\left\langle \dot{M}\right\rangle$;
$c$ is the speed of light.
Finally,
$W_{\rm Diss}$ in equation (\ref{thermal}) is the rate of the energy release
in the course of dissipative damping of the excited r mode; it is given by
\begin{equation}
W_{\rm Diss}= \frac{\widetilde{J}  M R^2  \Omega^2  \alpha^2}{ \tau_{\rm Diss}},
\label{Wdiss}
\end{equation}
where $M$ and $R$ are the NS mass and radius, respectively.
In all numerical calculations we adopt $M=1.4M_\odot$ and $R=10\,\rm km$.
$\Omega=2\pi \nu$ is the NS spin frequency (measured in $s^{-1}$);
$\alpha$ is the dimensionless amplitude of the r mode
(for definition of $\alpha$ see, e.g.,
formula 1 of \citealt{gck14_large});
$\tau_{\rm Diss}$ is the r-mode damping timescale
due to various dissipative processes
(shear and bulk viscosities, Ekman layer dissipation, mutual friction etc., see \citealt{ak01}).
In the range of temperatures relevant to LMXB systems
the most powerful mechanisms of r-mode dissipation
not too close to the stability peak
are the shear viscosity and dissipation in the Ekman layer.
\footnote{Here we do not discuss any `exotic' 
dissipation processes such as bulk viscosity due to hyperons or quarks.}
Due to large uncertainties inherent to Ekman-layer calculations
(see, e.g., \citealt{lu01, yl01, ak01, Rieutord01,Rieutord01_erratum,mendell01,km03,ga06a})
we, for definiteness, consider
shear viscosity (to be more precise, electron shear viscosity)
as the only dissipation mechanism
for r-modes far from the resonances with superfluid inertial modes
(see \citealt{gck14_large} for an expanded discussion and justification of our choice).
Note that near the resonances (on the slopes of the stability peaks)
superfluid inertial modes admix to the r-mode solution
(r-mode transforms into a superfluid inertial mode and vice versa) so that
dissipation there is mainly determined
by the mutual friction mechanism,
which is extremely efficient for superfluid modes (\citealt{lm00}).
Finally, $\widetilde{J}$ in equation (\ref{Wdiss})
is a numerical coefficient,
which equals $\widetilde{J}\approx 1.6353 \times 10^{-2}$
for an r mode with multipolarity $l=m=2$
and the polytropic equation of state (EOS)
with polytropic index $n = 1$
($P\propto \rho^{1+1/n}$,
where $P$ and $\rho$ are, respectively, the pressure and density of matter).

(ii) Evolution of stellar spin frequency $\Omega$,
\begin{equation}
\frac{d \Omega}{dt} = - \frac{2 \, Q \, \alpha^2 \, \Omega}{\tau_{\rm Diss}}+\dot{\Omega}_{\rm ext},
\label{dOmegadt}
\end{equation}
where the first term in the right-hand side represents
NS spin down due to $r$-mode dissipation;
$Q=l(l+1)\widetilde{J}/(4\widetilde{I})$;
$\widetilde{I}$ is defined as $I=\widetilde{I}MR^2$
($I$ is stellar moment of inertia).
For polytropic EOS with $n=1$
one has $\widetilde{I}\approx 0.261$;
$\dot{\Omega}_{\rm ext}$
in equation (\ref{dOmegadt}) is the rate of change of the NS spin frequency
due to
other spin down mechanisms, such as
accretion torque, magneto-dipole radiation, 
gravitational wave emission by possible mountains on an NS etc.

(iii) Evolution of $r$-mode amplitude $\alpha$
\begin{equation}
\frac{d \alpha}{dt} = -\alpha \left( \frac{1}{\tau_{\rm GR}}+\frac{1}{\tau_{\rm Diss}} \right),
\label{dadt}
\end{equation}
where $\tau_{\rm GR}$ is the gravitational radiation timescale,
which is negative, because emission of gravitational waves excites r mode.
For polytropic EOS with $n=1$
it can be calculated as (see \citealt{ak01})
\begin{equation}
    \tau_{\rm GR} = -\tau_{\rm GR \, 0}
        \left(\frac{M}{1.4 M_\odot}\right)^{-1}
        \,\left(\frac{R}{10\,\mathrm{km}}\right)^{-2l}
        \,\left(\frac{\nu}{1 \mathrm{kHz}}\right)^{-2l-2},
\label{tauGR2}
\end{equation}
where $\tau_{{\rm GR} \, 0}\approx 46.4$~s
for $l=m=2$ $r$ mode.
Instability window is given by the condition
$|\tau_{\rm GR}|<\tau_{\rm Diss}$,
which means that excitation of r modes occurs faster than their damping.
The boundary of the instability window defines
the instability curve at which $|\tau_{\rm GR}|=\tau_{\rm Diss}$.
Note, that we
kept only the leading order terms in $\alpha$ in Equations (\ref{Wdiss})--(\ref{dadt}),
assuming that $\alpha \ll 1$;
we also neglected the term
$\propto \dot{\Omega}_{\rm ext}/\Omega$ in Equation (\ref{dadt}),
because $\dot{\Omega}_{\rm ext}/\Omega \ll |1/\tau_{\rm GR}|$.
Moreover, Equations (\ref{thermal})--(\ref{dadt}) 
are only valid as soon as oscillations are linear, 
that is they are far from the saturation
defined either by non-linear coupling to other oscillation modes (see, e.g., \citealt*{btw07})
or by some other non-linear processes such as those discussed by \cite*{hga14} and \cite*{ahs15}.

These equations can be rewritten in a more compact form as
\begin{eqnarray}
\frac{dT^\infty}{dt}=F_1(T^\infty,\Omega)\Omega^2 A-F_2(T^\infty),\label{1}\\
\frac{d\Omega}{dt}=-G(T^\infty,\Omega) A \Omega+ \dot{\Omega}_{\rm ext},\label{2}\\
\frac{dA}{dt}=-A H(T^\infty,\Omega),
\label{3}
\end{eqnarray}
where $A\equiv \alpha^2$,
\begin{equation}
F_1(T^\infty,\Omega)\equiv \frac{\widetilde{J}  M R^2}{ C_{\rm tot} \tau_{\rm Diss}},\label{F1}
\end{equation}
\begin{equation}
F_2(T^\infty)\equiv \frac{L_{\rm cool} - K_{\rm n}  \dot{M} c^2}{ C_{\rm tot}},
\label{F2}
\end{equation}
\begin{equation}
G(T^\infty,\Omega)\equiv \frac{2 \, Q }{\tau_{\rm Diss}}, \label{Gdef}
\end{equation}
\begin{equation}
H(T^\infty,\Omega)\equiv 2\left( \frac{1}{\tau_{\rm GR}}+\frac{1}{\tau_{\rm Diss}} \right).
\end{equation}

\begin{figure}
    \begin{center}
        \leavevmode
        \epsfxsize=3.0in \epsfbox{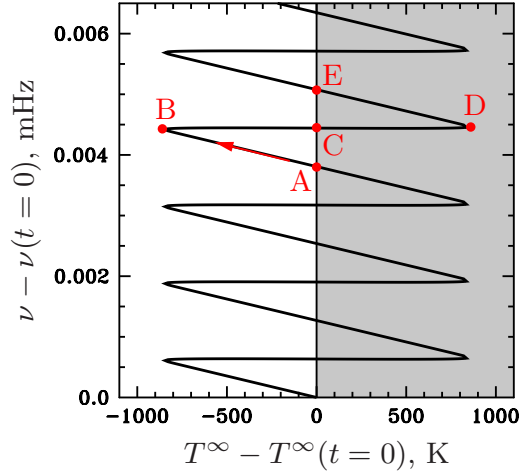}
    \end{center}
    \caption{Oscillations of the evolutionary track near the instability curve.
        Shown are variations of the rotation frequency (in millihertzs) and internal temperature (in Kelvins)
        starting from some initial moment of time $t=0$
                for the source 4U 1608-522. The time interval is $\sim 3\,\rm years$,
    direct Urca processes are forbidden, see Section \ref{Sec:obs} for details.}
    \label{Fig:evolution}
\end{figure}

Equations (\ref{1})--(\ref{3})
fully describe evolution of NSs in LMXBs.
In what follows we will use them
for the analysis of oscillations of NS evolution tracks
near the left edge of the stability peak, mentioned in Section \ref{Sec:Intro}.
An example of such oscillations is presented in
Fig.\ \ref{Fig:evolution},
which is a zoomed in fragment of the track
in the vicinity of
the stability peak,
analogous to that shown in panel (b) of Fig.\ \ref{Fig:inst1}.
The region filled with grey is the stability region inside the peak,
almost vertical straight line is the instability curve (the left edge of the peak).
Zigzag is the evolutionary track.
At point A of the cycle
the r mode amplitude has the lowest value which is not sufficient
to keep a star at the given temperature.
Thus the NS cools down, crosses the instability curve and enters the instability region
where r mode starts to grow.
As the r mode amplitude increases, the heating by the dissipation of the excited r mode
becomes more and more efficient
and slows down the NS cooling rate.
At some moment the NS reaches ``thermal balance'' point (point B in Fig.\ \ref{Fig:evolution}),
where heating by the accretion and r-mode dissipation exactly compensates NS cooling.
Further increase of the r-mode amplitude
leads to a temperature growth,
that brings the star to the instability curve.
After NS crosses the instability curve (point C),
the r-mode rise
replaces with the
r-mode decay, and NS heating reduces.
When a thermal balance point is reached again (now it is point D),
further decreasing of r-mode amplitude results in a net NS cooling.
As the NS cools down to the instability curve (point E), the cycle repeats.
In what follows,
to avoid confusion with the r-mode oscillations,
we will call these oscillations $\alpha$-oscillations
(emphasizing that it is the oscillations
of the r-mode amplitude $\alpha$).
In the next section $\alpha$-oscillations are described analytically.

\section{$\alpha$-oscillations near the edge of the stability peak}
\label{Sec:Equations}
%

In what follows, to simplify notations
the
redshifted internal stellar temperature $T^\infty$
will be denoted by $T$.
Unlike the oscillation amplitude squared, $A$, 
which varies significantly 
in the course of $\alpha$-oscillations,
the variations of $T$ and $\Omega$ (or $\nu$) are small.
Thus, for the analysis of $\alpha$-oscillations
we will present $T$ and $\Omega$
as a sum of the equilibrium solution
(see below)
which we will denote by the subscript 0,
and a small perturbation near this solution,
\begin{eqnarray}
T=T_0+\delta T, \label{pertT}\\
\Omega=\Omega_0+\delta \Omega.
\label{pertOmega}
\end{eqnarray}
The equilibrium solution assumes that $\Omega=\Omega_0$, $T=T_0$, and $A=A_0$
do not oscillate and the star track follows
the left edge of the stability peak,
which is defined by the condition
\begin{equation}
H(T_0,\Omega_0)= 0.
\label{stability_curve}
\end{equation}
It gives us the dependence $\Omega_0(T_0)$.
The time dependence of these quantities is driven by
the following equations (cf.\ equations\ \ref{1} and \ref{2})
\begin{eqnarray}
\frac{dT_0}{dt}=F_1(T_0,\Omega_0)\Omega_0^2 A_0-F_2(T_0), \label{T0temp}\\
\frac{d\Omega_0}{dt}=-G(T_0,\Omega_0) A_0 \Omega_0+ \dot{\Omega}_{\rm ext},
\label{Omega0temp}
\end{eqnarray}
hence
\begin{eqnarray}
\frac{\Omega_0}{T_0}\frac{dT_0}{d\Omega_0}=
-\frac{F_1(T_0,\Omega_0)\Omega_0^3 A_0-F_2(T_0)\Omega_0}{G(T_0,\Omega_0) A_0 \Omega_0 T_0- \dot{\Omega}_{\rm ext}T_0}.
\label{A0_eq}
\end{eqnarray}
This equation should be understood as an equation for $A_0$,
because $dT_0/d\Omega_0$ is known from Equation (\ref{stability_curve}).
Generally $(\Omega_0/T_0) dT_0/d\Omega_0 \lesssim 1$,
while the ratio of the
terms $F_1(T_0,\Omega_0) \Omega_0^3 A_0$ and $G(T_0,\Omega_0) A_0 \Omega_0 T_0$,
entering the right-hand-side of Equation (\ref{A0_eq}), can be estimated as
\begin{eqnarray}
\frac{F_1(T_0,\Omega_0) \Omega_0^2}{G(T_0,\Omega_0) T_0}=\frac{2I\Omega_0^2}{l(l+1)C_{\rm tot}T_0},
\label{estim}
\end{eqnarray}
that is it is of the order of the ratio of rotational energy to the thermal energy.
In the range of $\Omega_0$ and $T_0$ relevant to LMXBs,
this ratio is $\sim 10^5\gg 1$.
At the same time,
the external torque $\dot{\Omega}_{\rm ext}$ in Equation (\ref{A0_eq})
is generally
(at not too small $A_0$)
comparable to
the torque due to r modes, $G(T_0,\Omega_0) A_0 \Omega_0$.
This means that to satisfy Equation (\ref{A0_eq})
$A_0$, to a very good approximation,
should be given by the thermal equilibrium condition,
\begin{equation}
F_1(T_0,\Omega_0)\Omega_0^2 A_0\approx F_2(T_0).
\label{Tequil}
\end{equation}

 Substituting now Equations (\ref{pertT}) and
(\ref{pertOmega}) into (\ref{1})--(\ref{3}) one obtains the
following system of equations for r-mode oscillation
amplitude squared, $A$, and perturbations $\delta T$ and
$\delta \Omega$ (note that the quantity $A$ strongly
deviates from its equilibrium value $A_0$, while
perturbations $\delta T$ and $\delta \Omega$ are small
so that only first-order terms $\propto \delta T$ and $\delta \Omega$
are retained in these equations)
\begin{eqnarray}
\frac{d\delta T}{d t}=F_1(T_0,\Omega_0)\Omega_0^2 (A-A_0)+\underline{\underline{\left(\frac{\partial F_1(T,\Omega)}{\partial T}|_{T_0,\Omega_0}\Omega_0^2 A-\frac{\partial F_2(T)}{\partial T}|_{T_0}\right)\delta T}} \nonumber\\+\underline{\underline{\left(2 F_1(T_0,\Omega_0)\Omega_0 A+\frac{\partial F_1(T,\Omega)}{\partial \Omega}|_{T_0,\Omega_0}\Omega_0^2 A\right) \delta \Omega}},\label{dT}\\
\frac{d\delta \Omega}{d t}=-G(T_0,\Omega_0)\Omega_0(A-A_0)-\underline{\underline{\frac{\partial G(T,\Omega)}{\partial T}|_{T_0,\Omega_0}\Omega_0 A \delta T}} \nonumber\\-\underline{\underline{\left(\frac{\partial G(T,\Omega)}{\partial \Omega}|_{T_0,\Omega_0}\Omega_0 A +G(T_0,\Omega_0) A\right) \delta \Omega}},\label{dOmega}\\
\frac{dA}{d t}= -A\left(\frac{\partial H(T,\Omega)}{\partial T}|_{T_0,\Omega_0} \delta T+\frac{\partial H(T,\Omega)}{\partial \Omega}|_{T_0,\Omega_0} \delta \Omega \right). \label{dA}
\end{eqnarray}
This system does not depend on the external torque $\dot{\Omega}_{\rm ext}$ explicitly
(only by means of $A_0$, which is a function of $\dot{\Omega}_{\rm ext}$),
thus it is valid for both ascending-the-peak and descending-the-peak NSs,
and is even valid
in the absence of accretion
(that is, for millisecond pulsars,
if they are attached to a peak or for hot widows/HOFNARs, see \citealt*{cgk14}).
Generally, the underlined terms are small, because $T$ and $\Omega$
only slightly
deviate from their equilibrium values,
while, as we already mentioned above,
the deviation of r-mode amplitude squared, $A$, is not small.
Thus the leading terms in Equations (\ref{dT}) and (\ref{dOmega})
are those proportional to $A-A_0$.
Dividing now Equation (\ref{dT}) by Equation (\ref{dOmega}) and keeping the leading terms only, one can estimate
[(cf.\ Equation (\ref{estim})]
\begin{eqnarray}
\frac{\Omega_0}{T_0} \frac{\delta T}{\delta \Omega}
\sim \frac{F_1(T_0,\Omega_0) \Omega_0^2}{G(T_0,\Omega_0) T_0}
\sim 10^5\gg 1.
\end{eqnarray}
Thus it is easy to verify that the terms
in the system of Equations (\ref{dT})--(\ref{dA}) $\propto \delta \Omega$ are small and can be skipped
(then the equation for $\delta \Omega$ decouples and can be disregarded in what follows).
We are left then with
\begin{eqnarray}
\frac{{\rm d}\delta T}{{\rm d} t}=F_1(T_0,\Omega_0)\Omega_0^2 (A-A_0)+\left(\frac{\partial F_1(T,\Omega)}{\partial T}|_{T_0,\Omega_0}\Omega_0^2 A-\frac{\partial F_2(T)}{\partial T}|_{T_0}\right)\delta T, \label{dT1}
\\
\frac{{\rm d}A}{{\rm d} t}= -A\frac{\partial H(T,\Omega)}{\partial T}|_{T_0,\Omega_0} \delta T . \label{dAcut}
\end{eqnarray}
The time derivative of the Equation (\ref{dAcut}) is:
\begin{eqnarray}
\frac{\rm d}{{\rm d}t}\frac{{\rm d\, \log}A}{{\rm d} t}= -
\frac{\partial H(T,\Omega)}{\partial T}|_{T_0,\Omega_0}
\frac{{\rm d}\delta T}{{\rm d}t} -\frac{\rm d}{{\rm
d}t}\frac{\partial H(T,\Omega)}{\partial T}|_{T_0,\Omega_0}
\delta T . \label{dAcut2}
\end{eqnarray}
Substituting now ${\rm d} \delta T/{\rm d} t$ from Equation (\ref{dT1})
into (\ref{dAcut2}), we obtain
\begin{eqnarray}
\frac{{\rm d}^2y}{{\rm d}t^2}= -\beta(A-A_0)-
\frac{\partial H(T,\Omega)}{\partial T}|_{T_0,\Omega_0}\gamma (A)\delta T-\frac{\rm d}{{\rm d}t}\frac{\partial H(T,\Omega)}{\partial T}|_{T_0,\Omega_0} \delta T,
\label{oscill_y_diss00}
\end{eqnarray}
where we introduced a new variable $y=\log A$ and defined
\begin{eqnarray}
\beta \equiv \frac{\partial H(T,\Omega)}{\partial T}|_{T_0,\Omega_0} F_1(T_0,\Omega_0)\Omega_0^2,\\
\gamma (A) \equiv \frac{\partial F_1(T,\Omega)}{\partial T}|_{T_0,\Omega_0}\Omega_0^2 A-\frac{\partial F_2(T)}{\partial T}|_{T_0}.
\label{gamma}
\end{eqnarray}

The third term in the right-hand-side of equation (\ref{oscill_y_diss00}) is much smaller
than the second one,
because
$\partial H(T,\Omega)/\partial T|_{T_0,\Omega_0}$ evolves on a timescale of peak climbing (spin-up timescale), $\gtrsim 10^8$~yr, which is much longer than $\gamma^{-1} \sim t_{\rm{cool}}$,
where $t_{\rm cool}$ is the
cooling timescale, $t_{\rm{cool}}\approx T_0/F_2(T_0)\lesssim 10^6$~yr.
Thus, we will omit the third term in what follows,
\begin{eqnarray}
\frac{{\rm d}^2y}{{\rm d}t^2}= -\beta(A-A_0)-
\frac{\partial H(T,\Omega)}{\partial T}|_{T_0,\Omega_0}\gamma (A)\delta T.
\label{oscill_y_diss0}
\end{eqnarray}
Substituting now $\delta T$ from equation (\ref{dAcut}) into (\ref{oscill_y_diss0}), we arrive at
\begin{eqnarray}
\frac{{\rm d}^2y}{{\rm d}t^2}= -\beta({\rm e}^y- {\rm e}^{y_0})+ \gamma (A)\frac{{\rm d}y}{{\rm d}t},
\label{oscill_y_diss}
\end{eqnarray}
where $y_0 \equiv \log  A_0$.
Equation (\ref{oscill_y_diss}) describes
{\it nonlinear }
oscillations of the squared amplitude $A$
near the equilibrium point $A=A_0$ with dissipation/excitation described by the last term, which is typically small.

Note that the quantity $A$,
averaged over the period $\widehat{P}$ of $\alpha$-oscillations,
equals $\left\langle A \right\rangle=A_0$,
i.e., the equilibrium r-mode amplitude
$\alpha_0 \equiv \sqrt{A_0}$ is, at the same time, the root mean square of $\alpha$.
Indeed, the integral of the function $\delta T$ over $\widehat{P}$ should vanish
(we ignore a negligible change of the equilibrium temperature $T_0$ on a timescale of $\widehat{P}$),
\begin{eqnarray}
\int^{\widehat{P}} d(\delta T) = 0.
\end{eqnarray}
On the other hand
\begin{eqnarray}
\int^{\widehat{P}} d(\delta T)=\int^{\widehat{P}} \frac{d (\delta T)}{d t} dt=\int^{\widehat{P}} F_1(T_0,\Omega_0)\Omega_0^2 (A-A_0)  dt, \label{intdT}
\end{eqnarray}
where use has been made of (\ref{dT1}) with the last small term $\propto \delta T$ omitted.
Thus, because $T_0$ and $\Omega_0$ are
almost constants on a timescale of $\alpha$-oscillation period,
\begin{eqnarray}
\int^{\widehat{P}} A  dt =  A_0 \widehat{P}
\label{alpha_rms}
\end{eqnarray}
or $\left\langle A \right\rangle=A_0$.

To study
$\alpha$-oscillations driven by equation (\ref{oscill_y_diss}),
it is convenient to introduce a notion of the ``energy'' $\widehat{E}$
for this equation
[this function is conserved,
$\dot{\widehat{E}}=0$, when $\gamma=0$;
then it is just the first integral of
Equation (\ref{oscill_y_diss})],
\begin{eqnarray}
\widehat{E}\equiv \frac{1}{2}\dot{y}^2+\beta({\rm e}^y-y\, {\rm e}^{y_0}). \label{Energy}
\end{eqnarray}
Equation (\ref{oscill_y_diss}) is equivalent to the following formula,
\begin{eqnarray}
\dot{\widehat{E}}=\gamma (A) \dot{y}^2,
\end{eqnarray}
so that the energy variation averaged over the $\alpha$-oscillation period, $\widehat{P}$, is
\begin{eqnarray}
\left\langle \dot{\widehat{E}}\right\rangle=\left\langle \gamma (A) \dot{y}^2\right\rangle=\frac{1}{\widehat{P}} \int^{\widehat{P}} \gamma (A) \dot{y}^2 dt,
\label{int}
\end{eqnarray}
If $\left\langle \dot{\widehat{E}}\right\rangle$
is negative $\alpha$-oscillations are damped,
otherwise they are excited.
As long as dissipation/excitation is weak,
$y$ in this integral can be calculated from Equation (\ref{oscill_y_diss})
with $\gamma(A)=0$.
Equivalently, $y$ can be found from
Equations (\ref{dT1}) and (\ref{dAcut})
with the last term [$=\gamma (A) \delta T$]
in equation (\ref{dT1}) omitted.
To analyze integral in Equation (\ref{int}) let us notice that [see also (\ref{intdT})]
\begin{eqnarray}
0=\int^{\widehat{P}} d (\delta T^3)=\int^{\widehat{P}} \frac{d (\delta T^3)}{d t} dt=3\int^{\widehat{P}} \delta T^2 \frac{d (\delta T)}{d t} dt \nonumber
\\
=3\int^{\widehat{P}}
\left(\frac{\dot{y}}{\frac{\partial H(T,\Omega)}{\partial T}|_{T_0,\Omega_0}}\right)^2 F_1(T_0,\Omega_0)\Omega_0^2 (A-A_0)  dt,
\end{eqnarray}
where use has been made of Equations (\ref{dT1}) (with the last term neglected) and (\ref{dAcut}).
Since $T_0$ and $\Omega_0$ are
almost constants on a timescale of $\alpha$-oscillation period,
corresponding quantities can be
factored out of
the integral, yielding
\begin{eqnarray}
\int^{\widehat{P}} A \dot{y}^2 dt =  A_0 \int^{\widehat{P}} \dot{y}^2 dt.
\end{eqnarray}
Hence,
\begin{eqnarray}
\left\langle \dot{\widehat{E}}\right\rangle=\frac{1}{\widehat{P}} \int^{\widehat{P}} \gamma (A) \dot{y}^2 dt=\frac{1}{\widehat{P}} \int^{\widehat{P}} \left(\frac{\partial F_1}{\partial T}|_{T_0,\Omega_0}\Omega_0^2 A-\frac{\partial F_2}{\partial T}|_{T_0}\right) \dot{y}^2 dt\nonumber\\
=\frac{1}{\widehat{P}} \left(\frac{\partial F_1}{\partial T}|_{T_0,\Omega_0}\Omega_0^2 A_0-\frac{\partial F_2}{\partial T}|_{T_0}\right) \int^{\widehat{P}}  \dot{y}^2 dt=\frac{1}{\widehat{P}} \gamma (A_0) \int^{\widehat{P}}  \dot{y}^2 dt, \label{stability}
\end{eqnarray}
which means that when $\gamma (A_0)$ is negative
we have dissipation of $\alpha$-oscillations;
in the opposite case $\alpha$-oscillations excite.

Notice, that the value of $\gamma (A_0)$ depends on the geometry of the stability peak ---
for wider peaks the derivative $-\partial \tau_{\rm diss}/\partial T|_{T_0,\Omega_0}$, and hence $\partial F_1(T,\Omega)/\partial T|_{T_0,\Omega_0}$ [see Equation (\ref{F1})], is smaller. It results in decrease of $\gamma (A_0)$ with increasing peak width;
for sufficiently wide peaks $\gamma (A_0)$ is negative and $\alpha$-oscillations are damped.

The same criterion follows from the analysis
of the linear oscillations (when $A-A_0 \ll A_0$).
The stability of the solution to the system of linear equations
\begin{eqnarray}
\frac{d\delta T}{d t}=F_1(T_0,\Omega_0)\Omega_0^2 \delta A+\gamma (A_0) \delta T,\\
\frac{d \delta A}{d t}= -A_0 \frac{\partial H(T,\Omega)}{\partial T}|_{T_0,\Omega_0} \delta T
\end{eqnarray}
requires $\gamma (A_0)<0$ and $F_1(T_0,\Omega_0)\Omega_0^2 A_0
\frac{\partial H(T,\Omega)}{\partial T}|_{T_0,\Omega_0}>0$.
The last condition is satisfied automatically
at the left edge of the peak,
while the first one coincides with the stability condition
in the non-linear case.
This coincidence was not necessary:
A system, unstable in the linear approximation,
can, in principle, have stable non-linear solutions (and vice versa).
We can conclude that in the case when
\begin{equation}
\gamma (A_0)>0
\label{cond}
\end{equation}
a star cannot move along
the peak without $\alpha$-oscillations.
In all numerical examples considered below this inequality is satisfied.
We note that a similar criterion was obtained,
in the linear approximation, by \cite{whl01}.
This reference analyzed stability of an NS track in the vicinity of the instability curve
(positive, negative and horizontal slopes of the instability curve were discussed).

The above consideration allowed $\alpha$ to vary without
limits. However, the r-mode oscillation amplitude cannot
\textit{increase} infinitely. It is
limited by the saturation amplitude $\alpha_{\rm sat}$,
which is determined by a non-linear interaction of r mode with other oscillation modes,
or some other nonlinear processes such as those discussed by \cite{hga14} and \cite{ahs15}%
\footnote{Note that the equation (\ref{oscill_y_diss}) is only
valid if $\alpha<\alpha_{\rm sat}$. 
For example,  if the saturation
is defined by the nonlinear mode coupling, 
this condition means that $\alpha$ should be lower
than the parametric instability thresholds for all
nonlinear mode couplings. For larger r-mode amplitudes the
nonlinear mode coupling leads to excitation of daughter
modes, and this excitation should be described properly
(see, e.g., \citealt{btw04b,btw07,btw09}). For
definiteness, we apply a simplified model, assuming that r
mode saturation fixes amplitude $\alpha$ (and thus $y$) at
a constant value $\alpha=\alpha_{\rm sat}$
($y=y_\mathrm{sat}$) (see, e.g., \citealt{gck14_large} for
more details). \label{footnote:satur}
 }.
The value of $\alpha_{\rm sat}$ is rather uncertain, but in
what follows we will assume, in accordance with
\cite{btw07}, that it is of the order of $10^{-4}$, i.e.,
we take $\alpha_{\rm sat}=10^{-4}$
($y_\mathrm{sat}=-18.4$). Similarly, the oscillation
amplitude $\alpha$ cannot {\it decrease} infinitely. The
minimum value of $\alpha$ is defined by some spontaneous
excitation mechanism, that can be either of thermal origin
or induced, e.g., by accretion. For example, for internal
temperature $T=10^8\,\rm K$ r-mode energy becomes
comparable to $k_{\rm B}T$ ($k_{\rm B}$ is the Boltzmann
constant) at the threshold value $\alpha_{\rm th} \sim
10^{-29}$. This value seems to be too small to be a real
lower limit for $\alpha$, thus, in what follows, we adopt
$\alpha_{\rm th} =\alpha^\mathrm{fid}_{\rm th} =
10^{-12}$ as a fiducial value for illustration of our
results. 
Qualitatively, our main results are not sensitive to
$\alpha_{\rm th}$ [see a discussion after Eq.\
(\ref{alpha1}) and the footnote \ref{footnote:scaling}].

Strictly speaking, the derivation of the
criterion (\ref{cond}) is valid only while $\alpha_{\rm
th}<\alpha<\alpha_{\rm sat}$ during the whole oscillation
period. When $\alpha$ in the course of oscillations reaches
the value of $\alpha_{\rm th}$ or $\alpha_{\rm sat}$, the
above derivation is not applicable.
Then
each $\alpha$-oscillation cycle
starts
with the same initial condition at the left
edge of the peak:
It can be either $\alpha=\alpha_{\rm th}$ (in this case
the star enters the instability window
after the cycle starts,
see point $A$ in Fig.\ \ref{Fig:evolution}) or
$\alpha=\alpha_{\rm sat}$ (in this case
the star penetrates the stability region inside the peak
after the cycle starts,
see
point $C$ in Fig.\ \ref{Fig:evolution});
in both cases the cycle ``does not remember'' the
prehistory (initial conditions are the same for any such
cycle). For illustration, let us consider a cycle with the
initial condition $\alpha=\alpha_{\rm th}$. In the
beginning of the cycle (point A in Fig.\
\ref{Fig:evolution}) the star is at the instability curve
and $y=y_{\rm th}$ [$y_{\rm th}=\log (\alpha_{\rm th}^2)$],
$\widehat{E}=\beta({\rm e}^{y_{\rm th}}-y_{\rm th}\, {\rm
e}^{y_0})$ ($\dot{y}=0$ at the instability curve). If we
forget for a moment about ``non-conservation'' of the
energy $\widehat{E}$ during the cycle, then
$\widehat{E}={\rm const}$, and $y$ can reach its initial
value $y_{\rm th}$ only when $\dot{y}$ will vanish again,
that is at the instability curve (point E in Fig.\
\ref{Fig:evolution}). The result of the energy injection
into $\alpha$-oscillations will be that $y$ will reach
again the initial value $y_{\rm th}$ not exactly at the
instability curve, as strict energy conservation would
imply, but a little bit earlier (the r-mode amplitude wants
to decrease further, hence increasing the range of $y$
variation, but $y$ is limited by $y_{\rm th}$). This is,
however, not important for the next cycle, which again
starts with $y=y_{\rm th}$ and will be exactly the same as
the previous one. A similar analysis can be carried out for
the situation when $y_{\rm sat}$ limits the amplitude
growth.
 Notice that, anyway, satisfaction of the condition
(\ref{cond}) implies that $\alpha$-oscillations are
inevitable,
and they will grow until they reach $\alpha_\mathrm{th}$ or $\alpha_\mathrm{sat}$.

\section{Some oscillation parameters}
\label{Sec:parameters}

In the course of fully developed $\alpha$-oscillations r mode amplitude $\alpha$
varies in some range $\alpha_{\rm min}<\alpha<\alpha_{\rm max}$,
with either $\alpha_{\rm min}=\alpha_{\rm th}$ ($\alpha_{\rm max}<\alpha_{\rm sat}$)
or $\alpha_{\rm max}=\alpha_{\rm sat}$ ($\alpha_{\rm min}>\alpha_{\rm th}$).
Typically, if $\alpha_{\rm sat}$ is not too low, the former scenario is realized.
Let us estimate the value of $\alpha_{\rm max}$ in this case.
The `energy' of oscillations (\ref{Energy})
in the beginning of the cycle, when $y=y_{\rm th}$
at the instability curve, equals
(note that at that moment ${\dot y}=0$)
\begin{equation}
\widehat{E}=\beta({\rm e}^{y_{\rm th}}-y_{\rm th}{\rm e}^{y_0}).
\label{Emin}
\end{equation}
Then NS
penetrates the instability region where $y$ increases,
r mode heats the star and eventually NS reaches the instability curve with
the maximum (over the period) value of $y$,
given by $y=y_{\rm max}$.
Correspondingly,
the energy can be written as
(again, we have $\dot{y}=0$ at the instability curve)
\begin{equation}
\widehat{E}=\beta({\rm e}^{y_{\rm max}}-y_{\rm max}{\rm e}^{y_0}). \label{Emax}
\end{equation}
As have already been discussed above,
the energy of oscillations
is conserved approximately on a timescale of $\alpha$-oscillation period
(i.e.\ dissipation/excitation of oscillations is small).
Hence one can equate Equations (\ref{Emin}) and (\ref{Emax})
and get, neglecting ${\rm e}^{y_{\rm th}}$ in comparison to other terms,
\begin{equation}
\alpha_{\rm max} \approx \sqrt{y_{\rm max}-y_{\rm th}}\, \alpha_0
 \equiv k\, \alpha_0.
\label{alpha1}
\end{equation}
(We recall that $y_{\rm max}={\log \, \alpha_{\rm
max}^2}$.) Since $\left|y_{\rm max}\right|\ll \left|y_{\rm
th}\right|$ the coefficient $k=\sqrt{y_{\rm max}-y_{\rm
th}}$
depends mostly on $\alpha_{\rm th}$
and is not sensitive to, e.g., peak and/or stellar
parameters. For the value $\alpha_{\rm th}=10^{-12}$,
adopted in this paper, the coefficient $k$ varies,
depending on $\alpha_0$, in the range $k\approx (5\div 6)$.
Lower value of $\alpha_{\rm th}$ would lead to a higher
difference between $\alpha_{\rm max}$ and $\alpha_0$; for
example, for $\alpha_{\rm th}=10^{-29}$ (unrealistically
small value) $k$ varies in the range $k\approx (10.3\div
10.7)$. On the opposite, higher value of $\alpha_{\rm th}$
would lead to a smaller difference between $\alpha_{\rm
max}$ and $\alpha_0$; for example, $\alpha_{\rm
th}=10^{-10}$ would give $k \approx (4\div 5)$.

In the case when $\alpha_{\rm sat}$, rather than $\alpha_{\rm th}$,
is reached in the course of $\alpha$-oscillations
[this situation takes place when $\alpha_{\rm max}$, given by Equation (\ref{alpha1}),
is greater than $\alpha_{\rm sat}$],
energy conservation equation,
\begin{equation}
{\rm e}^{y_{\rm min}}-y_{\rm min}{\rm e}^{y_0}={\rm e}^{y_{\rm sat}}-y_{\rm sat}{\rm e}^{y_0},
\label{Econs}
\end{equation}
would provide us with the value of $\alpha_{\rm min}>\alpha_{\rm th}$.
Neglecting the small term ${\rm e}^{y_{\rm min}}$ in (\ref{Econs}),
it gives
\begin{equation}
\alpha_{\rm min}\approx \alpha_{\rm sat}\,{\rm exp}\left(-\frac{\alpha_{\rm sat}^2}{2\alpha_0^2}\right).
\end{equation}
Here we will not discuss this situation in detail.

Let us now estimate the time, $\tau$, that an NS spends near $\alpha_{\rm max}<\alpha_{\rm sat}$.
For that we expand $y(t)$ in the vicinity of $y_{\rm max}$ (near the instability curve)
in the Taylor series,
\begin{equation}
y \approx y_{\rm max}+\frac{1}{2}\ddot{y}t^2
\label{expan}
\end{equation}
(we remind that $\dot{y}=0$ at the instability curve).
From Equation (\ref{expan}) it follows that
$\alpha$ will decrease by a factor of $e\approx 2.71$ in time $\tau$ given by
\begin{equation}
\tau\approx \sqrt{-\frac{4}{\ddot{y}}} \,\,.
\end{equation}
Using Equation (\ref{oscill_y_diss}) one can express $\ddot{y}$ as
$\ddot{y}\approx -\beta {\rm e}^{y_{\rm max}}$
(we neglect ${\rm e}^{y_{0}}$ in comparison with ${\rm e}^{y_{\rm max}}$
and ignore a small dissipative term $\propto \gamma(A)$ in that equation).
Employing then the condition of the approximate heat balance (\ref{Tequil}), we get
\begin{equation}
\tau\approx \frac{2}{\sqrt{\beta A_{\rm max}}}\approx \frac{2}{k\sqrt{\frac{\partial H(T,\Omega)}{\partial T}|_{T_0,\Omega_0} F_2(T_0)}}. \label{tau}
\end{equation}
Depending on the parameters, $\tau$ is, typically,
of the order of hours-months (see Tables \ref{Tab:param1}--\ref{Tab:param3}).

\begin{figure}
    \begin{center}
        \leavevmode
        \epsfxsize=4.0in \epsfbox{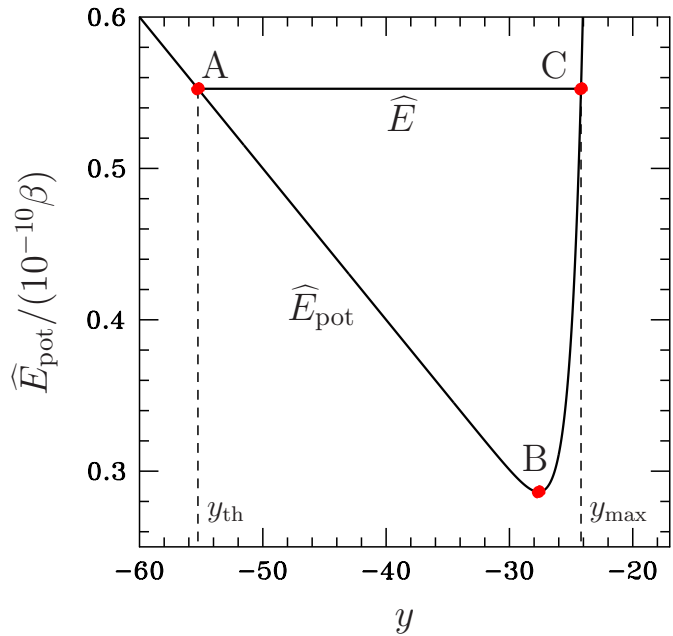}
    \end{center}
    \caption{The ``potential energy'' $\widehat{E}_{\rm pot}(y)$ normalized to $10^{-10}\beta$ as a function of $y$.
        The horizontal line shows the total energy $\widehat{E}/(10^{-10}\beta)$.
        The $\alpha$-oscillations proceed between points A and C.}
    \label{Fig:Epot}
\end{figure}

The period of $\alpha$-oscillations, $\widehat{P}$, is typically much longer than $\tau$.
It is easy to see it from the following rough estimate of the root-mean-square value of $\alpha$,
\begin{equation}
\alpha_0^2\sim \frac{\alpha_{\rm max}^2 \tau}{\widehat{P}},
\label{rough}
\end{equation}
where we assumed that $\alpha=\alpha_{\rm max}$ during the
time $\tau$, and the rest of the period $\alpha=0$.
Equation (\ref{rough}) gives
$\widehat{P} \sim k^2\tau$ [we
remind that $k\approx (5\div 6)$ for $\alpha_{\rm
th}=10^{-12}$]. Let us now estimate $\widehat{P}$ more
accurately. The energy (\ref{Energy}) can be interpreted as
a sum of ``kinetic energy'' $\dot{y}^2/2$ and ``potential
energy'',
\begin{equation}
\widehat{E}_{\rm pot}=\beta({\rm e}^y-y\, {\rm e}^{y_0}).
\label{pot_en}
\end{equation}
The dependence $\widehat{E}_{\rm pot}(y)$
is shown in Fig.\ \ref{Fig:Epot} (curve ABC). The straight horizontal line corresponds to the energy $\widehat{E}$ normalized to $10^{-10}\beta$. For this plot we used $\alpha_0=10^{-6}$ and  $\alpha_{\rm th}=10^{-12}$.
%
%
The evolution of $y$ and $\dot{y}$ in this potential is as follows.
At the time $t=0$ one has $y=y_{\rm th}$ and
 $\dot{y}=0$ (point A in Fig.\ \ref{Fig:evolution} and in Fig.\ \ref{Fig:Epot}).
Then $y$ increases, and $\dot{y}=0$ increases as well.
When $y$ is sufficiently small the evolution proceeds in practically linear potential
(stage A--B in Fig.\ \ref{Fig:Epot}) so that
\begin{equation}
\widehat{E}_{\rm pot}\approx -\beta y\, {\rm e}^{y_0}.
\label{pot_en_app}
\end{equation}
Then, eventually (around point B),
r mode amplitude excites for a time period of the order of $\tau$
(which is short in comparison to $\widehat{P}$),
and $\widehat{E}_{\rm pot}$ increases strongly,
because the first term in (\ref{pot_en}) becomes important (stage B--C).
Sharp rise of the potential energy (B--C stage) results in the ``velocity''
$\dot{y}$ decrease down to zero in point C
(corresponds to point C in Fig.\ \ref{Fig:evolution}).
After that $\dot{y}$ becomes negative, end evolution of $y$ follows in the opposite direction
(C--B--A stages in Fig.\ \ref{Fig:Epot},
which correspond to C--D--E stages in Fig.\ \ref{Fig:evolution}).
The short stages B--C and C--B are analogous to the elastic bounce of a ball from the wall.
Point B in Fig.\ \ref{Fig:Epot} corresponds to the minimum of the expression
(\ref{pot_en}) which is reached at $y=y_0$
(thus the point B of Fig.\ \ref{Fig:Epot} coincides with the points B and D in Fig.\ \ref{Fig:evolution}).
The potential energy drop at the stage A--B
can be estimated with the formula (\ref{pot_en_app}) as
\begin{equation}
\Delta \widehat{E}_{\rm pot}\approx -\beta \, {\rm e}^{y_0} (y_0-y_{\rm th}).
\label{D_pot_en}
\end{equation}
Correspondingly, the velocity at point B equals
\begin{equation}
\dot{y}_{\rm B}=\sqrt{2 \beta \, {\rm e}^{y_0} (y_0-y_{\rm th})},
\label{ydotB}
\end{equation}
where we made use of the energy conservation.
The evolution of $y$ at the stage $A-B$ is driven by the approximate equation
[we neglect the first term in Equation (\ref{pot_en}) at this stage],
\begin{eqnarray}
\frac{{\rm d}^2y}{{\rm d}t^2}= \beta {\rm e}^{y_0}.
\label{oscill_y_diss_app}
\end{eqnarray}
Integrating this equation, one gets
\begin{eqnarray}
\frac{{\rm d}y}{{\rm d}t}\approx \beta {\rm e}^{y_0}t+C
\label{oscill_y_diss_app_int}
\end{eqnarray}
with $C=0$ due to the initial condition $\dot{y}=0$.
Using Equations (\ref{ydotB})  and (\ref{oscill_y_diss_app_int}),
one can calculate the duration of the A--B stage,
\begin{equation}
t_{\rm AB}\approx\frac{\dot{y}_{\rm B}}{\beta {\rm
e}^{y_0}}\approx\sqrt{2(y_0-y_{\rm th})}\frac{\alpha_{\rm
max}}{\alpha_0}\frac{\tau}{2}=\sqrt{2(k^2-\log
\,k^2)}\,k\,\frac{\tau}{2},
\end{equation}
where the equality (\ref{tau}) has been used.
Then the period of $\alpha$-oscillations can be calculated as
\begin{equation}
\widehat{P}\approx 2 t_{\rm AB} \approx \sqrt{2(k^2-\log
\,k^2)}\,k\,\tau. \label{PPP}
\end{equation}
It is the sum of durations of A--B and B--A (after reflection in point C) stages.
The B--C and C--B stages proceed much faster
(the typical duration of these stages $\sim \tau \ll \widehat{P}$)
and the corresponding contributions to $\widehat{P}$ can be neglected.
Notice, that the ratio $\widehat{P}/\tau$ depends on $k$ only,
which is in turn mainly determined by $y_{\rm th}$,
and is not sensitive to the stellar and peak parameters.
The period $\widehat{P}$
varies in the range $\widehat{P}\approx (33 \div 48) \tau$
for $k=5\div 6$ (typical values for $\alpha_{\rm th}=10^{-12}$).
Thus, most of the time r mode is practically not excited
and increases up to $\alpha_{\rm max}$
for a short time interval of the order of $\tau$
($\sim 1/40$ of the oscillation period).

\section{Observational consequences}
\label{Sec:obs}

In the course of $\alpha$-oscillations
(when moving along the stability peak)
the temperature and spin frequency variations are very small, however,
two quantities vary significantly.
These are the r-mode amplitude, $\alpha$,
and spin frequency derivative, $\dot{\nu}$.
Variation of $\alpha$ results in modulation of gravitational radiation
while variation of $\dot{\nu}$
results in specific timing peculiarities of the signal from the star.
In this section we analyze
possible observational signatures
of such variations.

In Tables \ref{Tab:param1}--\ref{Tab:param3}
we introduce some parameters of $\alpha$-oscillations
for a number of NSs in LMXBs,
which climb a peak
within the scenario
of  \cite{gck14_short, gck14_large}.
Unfortunately, up to now there are no calculations
of temperature-dependent spectra of rotating superfluid NSs,
so currently we do not know the exact resonance temperatures $T_{\rm res}$ and the widths of the stability peaks,
characterized by the coupling parameter $s$
(this parameter describes coupling of r modes
with the superfluid inertial modes), see \cite{gck14_large} for more details.
Hence we will treat them as free parameters.
We have some idea about the possible typical values of $T_{\rm res}$ and $s$
from the calculations of temperature-dependent oscillation spectra
of non-rotating NSs (\citealt{cg11,kg11,gkcg13,gkgc14}).
Thus, in what follows we will choose them
in the reasonable range indicated by
these works.
For illustration,
we assume that each of the considered sources is attached to one of the four peaks,
which are characterized by $T_{\rm res}$ and $s$.
Two of these peaks are those described by \cite{gck14_large},
see figure 6 of that reference
(one is centered at $T_{\rm res}=4.5\times 10^7\,\rm K$
and is plotted for the coupling parameter $s=0.01$,
another is centered at $T_{\rm res}=1.5\times 10^8\,\rm K$
and has $s=0.001$),
the other two are centered at
$T_{\rm res}=7\times 10^7\,\rm K$
and $T_{\rm res}=10^8\,\rm K$
with the coupling parameters $s=0.01$ and $s=0.001$, respectively.
Notice that the lower temperature peaks have higher $s$,
in agreement with the results of \cite{gkcg13} and \cite{gkgc14}.
In our calculations
the spin frequencies
and surface effective temperatures for
the observed
sources
were taken from the
table I of \cite{gck14_large}
(if only an upper limit for the effective temperature is known,
we take the latter to be equal to this upper limit).

$\alpha$-oscillations depend on the NS cooling rate,
which is rather uncertain,
and can be enhanced strongly by the direct Urca (Durca) processes (see, e.g., \citealt{ykgh01}).
The most powerful of them is the neutron decay
into proton and lepton (electron or muon) with the emission of anti-neutrino,
$n \rightarrow p+l+\widetilde{\nu_{l}}$ ($l$ stands for a lepton),
and the corresponding inverse process,
$p+l \rightarrow n+\nu_{ l}$.
However, these are the threshold processes
which are forbidden at densities lower than some threshold densities.
These thresholds
are very sensitive to the EOS, and can be rather high.
For some EOSs these processes are always forbidden, even
for
the most massive NS configurations
(see, e.g., a recent model BSk19 by \citealt{pfcpg13}).
On the other hand, a majority of microscopic models predict that at a density
$\rho\sim (2-3)\rho_0$ ($\rho_0$ is the nuclear density) hyperons should appear
(see, e.g., \citealt{wcs12,bhzbm12,ghk14} and references therein).
In many of these models
the first hyperon species to appear
with growing density
is $\Lambda$ hyperons (see, e.g., \citealt{wcs12,bhzbm12,ghk14}).
Once $\Lambda$ hyperons appear the Durca processes
$\Lambda \rightarrow p+ l + \widetilde{\nu_l}$ and $p+ l  \rightarrow \Lambda+ \nu_l$
become possible.
Although NS cooling due to these processes is slightly less efficient (\citealt{pplp92,ykgh01})
than in the case of nucleonic ($npl$) Durca,
these are {\it not} the threshold processes,
and it is quite possible that
they operate in NSs with lower masses
than $npl$ Durca processes
(but the NS mass should be sufficiently high to host $\Lambda$ hyperons in their cores).
Both nucleonic and $\Lambda$ hyperonic Durcas
can be suppressed by baryon
(i.e., proton, $\Lambda$ hyperon, or neutron) superfluidity
whose properties are very uncertain.
However, it is very probable that protons are non-superfluid
at high densities (see, e.g., \citealt{plps09}),
while critical temperatures of $\Lambda$ hyperons are likely to be very low (\citealt{tmc03,tnyt06,ws10}),
so that $\Lambda$-hyperonic Durca process is only weakly suppressed (or completely unsuppressed).
On the other hand, the neutron superfluidity still can affect $npl$ Durca.
Neglecting possible suppression by superfluidity,
the temperature dependence of the emission rate for both  nucleonic
and $\Lambda$ hyperonic Durcas is the same,
so that the emissivity due to $\Lambda$ hyperonic Durca
is equivalent to the emissivity due to $npl$ Durca
from the smaller stellar volume
(namely, the radius of the central region of an NS,
where Durca processes operate, $R_{\rm D}$, should be smaller
for $npl$ Durca by approximately a factor of
$(m_\Lambda^\star r_{\rm \Lambda p}/m_{\rm n}^\star)^{1/3}\approx 0.36$,
see Appendix \ref{Sec:Appendix}).
Thus, just to illustrate the effect of an enhanced NS cooling on the oscillations,
we considered three cases:
($i$) all Durca processes are forbidden (Table \ref{Tab:param1});
($ii$) unsuppressed $\Lambda$ hyperonic Durca processes
operate in the central region of an NS with the radius $R_{\rm D}=0.1R$ (Table \ref{Tab:param2});
($iii$) the same as ($ii$) but for $R_{\rm D}=0.3R$ (Table \ref{Tab:param3}).
For more details on the Durca processes with $\Lambda$ hyperons see Appendix \ref{Sec:Appendix}.

Other parameters adopted in our calculations are presented in Table \ref{Tab:input}.
These are the spin frequency of an NS (see the first column and \citealt{gck14_large} for details),
accretion rate, averaged over quiescent and active states,
$ \dot{M}_{-10}$, in units of $10^{-10}\,\rm M_\odot/yr$
(see the second column and the corresponding references),
and the distance to the source, $d$, in $\rm kpc$ (the third column followed by the references).
We also show the resonance temperature ($T_{8\,\rm res}$, in units of $10^8\,\rm K$)
of the peak to which a given NS is assumed to be attached
and the coupling parameter $s$
corresponding to a given peak
(the last two columns).

\begin{table}
\caption{NS parameters.}
\begin{tabular}{l   c  c  l  c  l   c   c }
\hline
   Source                &  $\;\nu, \;\rm Hz \;$
     &  $\; \dot{M}_{-10}\;$ &
   & $\displaystyle \;d,\,\rm kpc\;$ &
   & $\displaystyle \;T_{8\,\rm res}\;$
   & $\displaystyle \;s$
\\
   \hline\hline
   4U 1608-522 & 620 & $3.6$ & \cite{hjwt07} & $4.1$ & \cite{Watts_et_al_08}   &    $1.5$  &   $0.001$    \\

   SAX J1750.8-2900 & 601  & $2.0$ & \cite{lowell_et_al_12} & $6.79$ & \cite{Watts_et_al_08} &    $1.5$    &   $0.001$       \\

   EXO 0748-676
             & 552   & $2.0$& $^a$  & $7.4$ & \cite{Watts_et_al_08}      &    $1.5$    &   $0.001$      \\

   Aql X-1   & 550    & $4.0$ & \cite{hjwt07} & $4.55$ & \cite{Watts_et_al_08}    &    $1.5$    &   $0.001$         \\

   SWIFT J1749.4-2807
     &518 & $2.0$ &$^b$ &$6.7$ & \cite{Wijnands_et_al_09} &    $1.5$    &   $0.001$        \\

   SAX J1748.9-2021 & 442 & $1.8$ & \cite{hjwt07} & $8.1$ & \cite{Watts_et_al_08}  &    $1.0$    &   $0.001$     \\

   IGR J17498-2921
    & 401 & $2.0$ &$^b$ &$7.6$ & \cite{Linares_et_al_11} &    $1.0$    &   $0.001$          \\

   KS 1731-260
   & 524 & $1.5$ &$^c$& $7.2$ & \cite{Watts_et_al_08}      & $0.7$ &  $0.01$  \\

      IGR J00291-5934
          & 599  & $0.05$ &$^d$ &$5$ & \cite{Watts_et_al_08}  &    $0.45$    &   $0.01$    \\

   MXB 1659-298 & 567 & $1.7$ & \cite{hjwt07} & $12$ & \cite{Watts_et_al_08} &    $0.45$  &    $0.01$         \\

   XTE J1751-305  & 435   & $0.06$ & \cite{heinke_et_al_09} & $9$ & \cite{Watts_et_al_08} & $0.45$ &  $0.01$  \\

             \hline
\end{tabular}
\\ $^a$ Accretion rate is unknown (see, however, \citealt{Degenaar_et_al_14}). We adopt here the value $2.0\times 10^{-10}\, \rm M_\odot/yr$.
\\ $^b$ Accretion rate is unknown. We adopt here the value $2.0\times 10^{-10}\, \rm M_\odot/yr$.
\\ $^c$ Only an upper limit, $1.5\times 10^{-9}\, \rm M_\odot/yr$  (accretion rate in active state), is known (\citealt{hjwt07}). We adopt here the value $1.5\times 10^{-10}\, \rm M_\odot/yr$.
\\ $^d$ We adopt here the value $5 \times 10^{-12}\, \rm M_\odot/yr$, somewhat in between the accretion rate from the paper by \cite{heinke_et_al_09} ($2.5 \times 10^{-12}\, \rm M_\odot/yr$, which is too low to balance even the photon cooling from the NS surface) and the accretion rate from the paper by \cite{Patruno10}, $(7-8) \times 10^{-12}\, \rm M_\odot/yr$.
 \label{Tab:input}
\end{table}

In Tables \ref{Tab:param1}--\ref{Tab:param3}
\footnote{Tables \ref{Tab:param1}--\ref{Tab:param3}
correspond to fiducial value
$\alpha_\mathrm{th}=\alpha^\mathrm{fid}_\mathrm{th}=10^{-12}$.
Equations (\ref{alpha1}), (\ref{tau}), and (\ref{PPP}) allow one to
rescale them to an arbitrary $\alpha_\mathrm{th}$
according to the formulas
 $\tau\approx \tau^\mathrm{fid}\, k^\mathrm{fid}/k$,
 $\widehat{P}\approx \widehat {P}^\mathrm{fid}\, k/k^\mathrm{fid}$,
 $|\Delta \nu|\approx|\Delta \nu^\mathrm{fid}|\, k/k^\mathrm{fid}$,
 $\sqrt{\widehat {P}}\langle h_0\rangle\approx \sqrt{\widehat {P}^\mathrm{fid}}\langle h_0\rangle\, \sqrt{k/k^\mathrm{fid}} $,
 and
 $\dot \nu_\mathrm{max}=\langle \dot \nu_\mathrm{acc}\rangle
 +k^2\, (\langle \dot \nu\rangle-\langle \dot \nu_\mathrm{acc}\rangle )$.
Here $\tau^\mathrm{fid}$,
 $\widehat {P}^\mathrm{fid}$,
 $|\Delta \nu^\mathrm{fid}|$, and
 $\sqrt{\widehat {P}^\mathrm{fid}}\langle h_0\rangle $ are
 the values from Tables \ref{Tab:param1}--\ref{Tab:param3},
 $k\approx \sqrt{2\,[\log (5\,\alpha_0)-\log \alpha_\mathrm{th}]}$
 and
 $k^\mathrm{fid}\approx \sqrt{2\,[\log (5\,\alpha_0)-\log  \alpha^\mathrm{fid}_\mathrm{th}]}$.
 This scaling is applicable if
 $\alpha_\mathrm{max}\approx k \alpha_0\le \alpha_\mathrm{sat}$.
The amplitude
$\alpha_0$ and spin frequency derivatives
 $\langle \dot \nu_\mathrm{acc}\rangle$
and
 $\langle \dot \nu\rangle$
do not depend on $\alpha_\mathrm{th}$.
\label{footnote:scaling}}
the first column presents $\alpha_0$ multiplied by $10^7$,
the second column shows how long an NS spends near the
maximum value of $\alpha$ during each oscillation period
[this is the parameter $\tau$ in days, calculated with the
formula (\ref{tau})], the third column contains periods of
$\alpha$-oscillations $\widehat{P}$ in years, given by
Equation (\ref{PPP}). The next four columns describe NS
timing behavior and are discussed in Section
\ref{subSec:timing}. The last column in Tables
\ref{Tab:param1}--\ref{Tab:param3} characterizes the
efficiency of gravitational wave emission, which is
discussed in Section \ref{subSec:gravrad}.

\begin{table}
\caption{Parameters of NS oscillations near the peak. No direct Urca.}
\begin{tabular}{l c c c c c c c c}
\hline
   Source               &  $\alpha_0$
   & $\displaystyle \tau$
   & $\displaystyle \widehat{P}$
   & $\displaystyle \left\langle \dot{\nu}_{\rm acc} \right\rangle$
   & $\displaystyle \dot{\nu}_{\rm max}$
   & $\displaystyle \left\langle \dot{\nu} \right\rangle$
   & $\displaystyle |\Delta \nu|$
     & $\displaystyle \sqrt{\widehat{P}}\,\langle h_0\rangle $
\\
   $ $        &  $\times 10^{7}$
   & $\displaystyle  $
   & $\displaystyle  $
   & $\displaystyle  \times 10^{14}$
   & $\displaystyle  \times 10^{14}$
   & $\displaystyle  \times 10^{14}$
   & $\displaystyle  \times 10^{6}$
   & $\displaystyle \times 10^{24}$
\\
   $ $              &  $ $
   & $\displaystyle [{\rm days}]$
   & $\displaystyle [{\rm yrs}]$
   & $\displaystyle [{\rm Hz\, sec^{-1}}]$
   & $\displaystyle [{\rm Hz\, sec^{-1}}]$
   & $\displaystyle [{\rm Hz\, sec^{-1}}]$
   & $\displaystyle [{\rm Hz}]$
     & $\displaystyle [{\rm Hz}^{-1/2}]$
\\
   \hline\hline

   4U 1608-522  &   $1.6$  &   $5.8$  & $0.57$ &  $6.8$& $-3.0$ & $6.4$   &$0.049$ & $2.1$ \\

   SAX J1750.8-2900   &   $2.3$    &   $5.1$        & $0.52$          & $3.8$& $-13$ & $3.2$ & $0.074$ & $1.6$\\

   EXO 0748-676 &    $2.8$  &   $8.2$       & $0.85$ &   $3.8$ & $-11$ &  $3.3$  & $0.10$ & $1.8$ \\

   Aql X-1     &    $2.2$    &   $11$       & $1.1$  &     $7.5$& $-0.66$ & $7.2$  & $0.078$ & $2.5$\\

   SWIFT J1749.4-2807   &    $3.6$    &   $11$       & $1.1$   & $3.8$ & $-11.3$ & $3.2$ & $0.14$ & $2.4$ \\

   SAX J1748.9-2021    &    $2.7$    &   $36$       & $3.7$  & $3.4$ & $0.55$ &  $3.3$& $0.088$ & $1.7$\\

   IGR J17498-2921      &    $3.0$    &   $72$       & $7.5$   & $3.8$ & $2.0$ &  $3.7$ & $0.11$ & $2.1$\\

   KS 1731-260 $^a$
        & $-$ &  $-$ & $-$ & $-$ & $-$ & $-$ & $-$ & $-$\\

   IGR J00291-5934    &    $0.12$    &   $103$  & $8.1$    & $0.094$ & $0.058$  & $0.092$ & $0.0032$ & $0.45$\\

   MXB 1659-298 $^a$
    &  $-$  & $-$ & $-$& $-$ & $-$ & $-$ & $-$ & $-$\\

   XTE J1751-305 $^b$
    & $-$ &  $-$ & $-$ & $-$ & $-$ & $-$ & $-$ & $-$\\

             \hline
\end{tabular}
\\ $^a$ The equilibrium temperature is higher than the observed one. Direct Urca process is required.
\\ $^b$ Equilibrium temperature is ``inside'' the low-temperature peak, thus this NS spins up in the stability region.

 \label{Tab:param1}
\end{table}

\begin{table}
\caption{Parameters of NS oscillations near the peak. Direct Urca in $0.1R$.}
\begin{tabular}{l c c c c c c c c}
\hline
   Source               &  $\alpha_0$
   & $\displaystyle \tau$
   & $\displaystyle \widehat{P}$
   & $\displaystyle \left\langle \dot{\nu}_{\rm acc} \right\rangle$
   & $\displaystyle \dot{\nu}_{\rm max}$
   & $\displaystyle \left\langle \dot{\nu} \right\rangle$
   & $\displaystyle |\Delta \nu|$
     & $\displaystyle \sqrt{\widehat{P}}\,\langle h_0\rangle $
\\
   $ $        &  $\times 10^{7}$
   & $\displaystyle  $
   & $\displaystyle  $
   & $\displaystyle  \times 10^{14}$
   & $\displaystyle  \times 10^{14}$
   & $\displaystyle  \times 10^{14}$
   & $\displaystyle  \times 10^{6}$
   & $\displaystyle \times 10^{24}$
\\
   $ $              &  $ $
   & $\displaystyle [{\rm days}]$
   & $\displaystyle [{\rm yrs}]$
   & $\displaystyle [{\rm Hz\, sec^{-1}}]$
   & $\displaystyle [{\rm Hz\, sec^{-1}}]$
   & $\displaystyle [{\rm Hz\, sec^{-1}}]$
   & $\displaystyle [{\rm Hz}]$
     & $\displaystyle [{\rm Hz}^{-1/2}]$
\\
   \hline\hline

   4U 1608-522    &    $25$    &   $0.33 $    & $0.040$ & $6.8$& $-3.0\times 10^{3}$ &   $-83$ & $0.85$ & $8.8$\\

   SAX J1750.8-2900   &    $28$   &   $0.37 $   & $0.046$   & $3.8$& $-3.1\times 10^{3}$ &   $-89$ & $1.0$ & $5.8$ \\

   EXO 0748-676    &    $40$  &   $0.53 $   & $0.066$   & $3.8$ & $-3.4\times 10^{3}$  &   $-96$ & $1.6$ & $7.0$ \\

   Aql X-1     &    $40$    &   $0.54$     & $0.067$   & $7.5$& $-3.4\times 10^{3}$ &  $-92$ & $1.6$ & $11$ \\

   SWIFT J1749.4-2807   &    $51$  &  $0.70$  & $0.088$   & $3.8$ & $-3.6\times 10^{3}$  &  $-102$ & $2.2$ & $9.4$\\

   SAX J1748.9-2021    &   $28$    &   $3.2$   & $0.39$   & $3.4$ & $-3.5\times10^{2}$ &  $-7.2$ & $0.98$ & $5.6$\\

   IGR J17498-2921   &   $41$  &   $4.9$   & $0.61$   & $3.8$ & $-3.8\times 10^{2}$   &    $-7.7$ & $1.7$ & $8.2$ \\

   KS 1731-260   &  $3.6$  &   $14$       & $1.5$         & $2.8$ & $-14$  &    $2.6$ & $0.20$ & $2.6$ \\

   IGR J00291-5934  &    $0.67$   &   $17$ & $1.6$       & $0.094$ & $-1.2$  &    $0.043$  & $0.019$ & $1.1$\\

   MXB 1659-298 $^a$
    & $-$ &  $-$  & $-$ &    $-$    & $-$& $-$ & $-$ & $-$\\

   XTE J1751-305     &    $1.8$    &   $99$  & $9.9$   & $0.11$ & $-0.91$   &    $0.076$  & $0.087$ & $1.5$\\

             \hline
\end{tabular}
\\ $^a$ The equilibrium temperature is higher than the observed one. Thus Durca in $0.1 R$ is not sufficient for this NS.

 \label{Tab:param2}
\end{table}

\begin{table}
\caption{Parameters of NS oscillations near the peak. Direct Urca in $0.3R$.}
\begin{tabular}{l c c c c c c c c}
\hline
   Source               &  $\alpha_0$
   & $\displaystyle \tau$
   & $\displaystyle \widehat{P}$
   & $\displaystyle \left\langle \dot{\nu}_{\rm acc} \right\rangle$
   & $\displaystyle \dot{\nu}_{\rm max}$
   & $\displaystyle \left\langle \dot{\nu} \right\rangle$
   & $\displaystyle |\Delta \nu|$
     & $\displaystyle \sqrt{\widehat{P}}\,\langle h_0\rangle$
\\
   $ $        &  $\times 10^{7}$
   & $\displaystyle  $
   & $\displaystyle  $
   & $\displaystyle  \times 10^{14}$
   & $\displaystyle  \times 10^{14}$
   & $\displaystyle  \times 10^{14}$
   & $\displaystyle  \times 10^{6}$
   & $\displaystyle \times 10^{24}$
\\
   $ $              &  $ $
   & $\displaystyle [{\rm days}]$
   & $\displaystyle [{\rm yrs}]$
   & $\displaystyle [{\rm Hz\, sec^{-1}}]$
   & $\displaystyle [{\rm Hz\, sec^{-1}}]$
   & $\displaystyle [{\rm Hz\, sec^{-1}}]$
   & $\displaystyle [{\rm Hz}]$
     & $\displaystyle [{\rm Hz}^{-1/2}]$
\\
   \hline\hline
   4U 1608-522 & $131$ & $0.061 $ & $8.1\times 10^{-3}$ & $6.8$ & $-8.8\times 10^{4}$ & $\,\,\,\, -2.4\times 10^{3}$& $4.6$ & $20$\\

   SAX J1750.8-2900 & $148$ & $0.069 $ & $9.3\times 10^{-3}$ & $3.8$& $-9.2\times 10^{4}$ & $\,\,\,\, -2.5\times 10^{3}$ & $5.4$ & $14$\\

   EXO 0748-676 & $206$ & $0.098$ & $0.013$  & $3.8$ & $-1.0\times 10^{5}$ & $\,\,\,\, -2.7\times 10^{3}$ & $8.5$ & $16$\\

   Aql X-1 & $209$ & $0.10 $ & $0.014$ & $7.5$ & $-1.0\times 10^{5}$ & $\,\,\,\, -2.7\times 10^{3}$ & $8.7$ & $27$\\

   SWIFT J1749.4-2807 & $265$ & $0.13 $ & $0.018$ & $3.8$ & $-1.1\times 10^{5}$ & $\,\,\,\,-2.8\times 10^{3}$ & $12$ & $22$\\

   SAX J1748.9-2021 & $146$ & $0.59 $ & $0.079$ & $3.4$ & $-1.0\times 10^{4}$  & $-2.8\times 10^2$ & $5.3$ & $13$\\

   IGR J17498-2921 & $213$  &   $0.91 $ & $0.12$ & $3.8$ & $-1.1\times 10^{4}$  &  $-3.0\times 10^2$ & $9.0$ & $19$\\

   KS 1731-260 &  $20$  &   $2.3$ & $0.28$  & $2.8$ & $-6.0\times 10^{2}$ &  $-15$ & $1.2$ & $6.5$\\

   IGR J00291-5934  & $3.4$ &  $3.1$ & $0.33$   & $0.094$ & $-38$ & $-1.2$ & $0.10$ & $2.6$\\

   MXB 1659-298  &  $3.9$  &   $4.4$ & $0.46$  & $3.2$ & $-30$ &  $2.0$ & $0.13$ & $1.2$\\

   XTE J1751-305 & $9.3$  & $18$ & $2.0$   & $0.11$ & $-32$ &  $-0.92$ & $0.49$ & $3.6$\\

             \hline
\end{tabular}

 \label{Tab:param3}
\end{table}

One can see that, as was stated in the previous section,
$\tau$ is indeed of the order of hours-months
and is larger for low-temperature sources.
This is so
because of two reasons.
First, $\tau$ roughly scales as $\sqrt{s}$,
being thus higher by a factor of three for two low-temperature peaks which have $s=0.01$.
\footnote{To show it we present $\partial H(T,\Omega)/\partial T|_{T_0,\Omega_0}$
entering the expression for $\tau$ (\ref{tau}) as $\partial H(T,\Omega)/\partial T|_{T_0,\Omega_0}
\approx 8 (s T_{\rm res})^{-1} |\tau_{\rm GR}|^{-1}(\tau_{\rm MF}/|\tau_{\rm GR}|)^{1/2}\propto s^{-1}$,
where $\tau_{\rm MF}$ is the mutual friction damping time of the superfluid mode.
To derive this expression we used the formulas of Section IVB
of \citealt{gck14_large} and considered a limiting case $x\rightarrow -\infty$.
}
Another factor further increasing this difference
is the fact that at higher temperatures
the function $F_2(T)=\left(L_{\rm cool} - K_{\rm n}  \dot{M} c^2\right)/ C_{\rm tot}$
is generally larger
at comparable accretion rates, while $\tau\propto 1/F_2(T_0)^{1/2}$.
The period $\widehat{P}\propto \tau$ and hence behaves similarly.

\subsection{Timing}
\label{subSec:timing}
Typical timing behaviour of an NS climbing a peak is the following.
Most of the time the stellar spin frequency and its derivative
are not affected by the r modes and are determined
by the external torque (accretion and spin-down torques due to, e.g., magneto-dipole losses).
Spin frequency derivatives at this stage,
averaged over quiescent and active phases,
$\left\langle \dot{\nu}_{\rm acc} \right\rangle$,
are presented in the fourth
column
of Tables \ref{Tab:param1}--\ref{Tab:param3}.
To evaluate $\left\langle \dot{\nu}_{\rm acc} \right\rangle$
we (for definiteness) neglected magneto-dipole losses and used the following simplified relation between $\left\langle\dot{\nu}_{\rm acc}\right\rangle$ and $\dot{M}$,
valid for non-magnetized NSs,
\begin{equation}
\left\langle\dot{\nu}_{\rm acc}\right\rangle =\dot{M}\, \frac{\sqrt{G  M  R}}{2\pi I}.
\label{nuacc}
\end{equation}
When an NS gets into the instability region the r-mode amplitude starts to grow
and soon approaches its maximum value.
NS affected by r-mode torque spins down rapidly at this stage.
The fifth column
in
the Tables
presents
the value of the spin frequency derivative
when r-mode amplitude is in its maximum, $\dot{\nu}_{\rm max}$.
This rapid spin down lasts
until
$\alpha$ is high, that is during the period of time $\tau$.
Finally, the sixth
column
contains the spin frequency derivative $\left\langle \dot{\nu}\right \rangle$,
averaged over the large number of $\alpha$-oscillation periods.

In observations a sudden spin down due to the action
of the r-mode torque, lasting the time $\tau$,
can be interpreted as an anti-glitch.
The sizes of the ``anti-glitches'' can be estimated as $|\Delta \nu|\approx (\left\langle \dot{\nu}_{\rm acc} \right\rangle-\dot{\nu}_{\rm max}) \tau$;
they are presented in the seventh column of the Tables.
Using Equations (\ref{Wdiss}), (\ref{dOmegadt}), (\ref{Tequil}), and (\ref{rough})
it is easy to show that $|\Delta \nu|$ can be expressed as
$|\Delta \nu|\approx 3 (L_\mathrm{cool}-K_{\rm n}  \dot M c^2) \widehat P/(2\pi I\Omega)$.
This formula
implies that the rotation energy losses
during an ``anti-glitch'' are three times larger than
the energy emitted by an NS during the whole $\alpha$-oscillation period, $\widehat P$.
As far as r-modes do not affect the spin frequency evolution
in the remaining time, the average spin down rate
is the same as in the absence of $\alpha$-oscillations
[see Equations (\ref{T0temp}), (\ref{Omega0temp}) and, e.g., equation (2) of \citealt{cgk14}];
thus the main effect of $\alpha$-oscillations
on the timing is
concentration
of all r-mode spin down
in
narrow time intervals $\sim \tau$.
Typically, these ``anti-glitches'' are rather small
(the values from the Tables lie in the interval $5.3\times 10^{-12}<|\Delta \nu/\nu| < 2.3\times 10^{-8}$)
and can contribute to the timing noise.

\begin{figure}
    \begin{center}
        \leavevmode
        \epsfxsize=6in \epsfbox{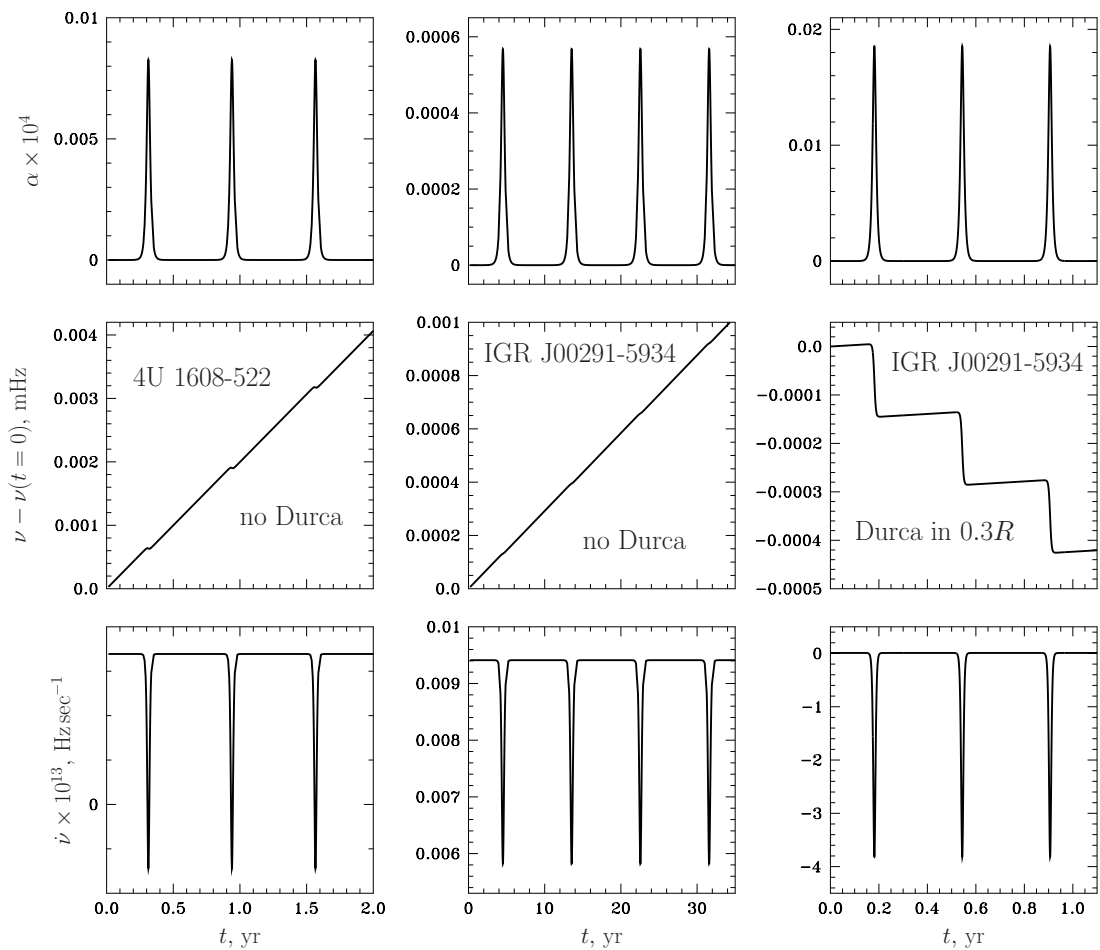}
    \end{center}
    \caption{R-mode amplitude (top panels), spin frequency (middle panels), and spin-down rate (bottom panels) as functions of time for (from left to right):
    (a) 4U 1608-522 without Durca, (b)
    IGR J00291-5934 without Durca, (c)
    IGR J00291-5934 with $\Lambda$-hyperonic Durca allowed for $r \leq 0.3R$.}
    \label{Fig:spin}
\end{figure}

\begin{figure}
    \begin{center}
        \leavevmode
        \epsfxsize=6.0in \epsfbox{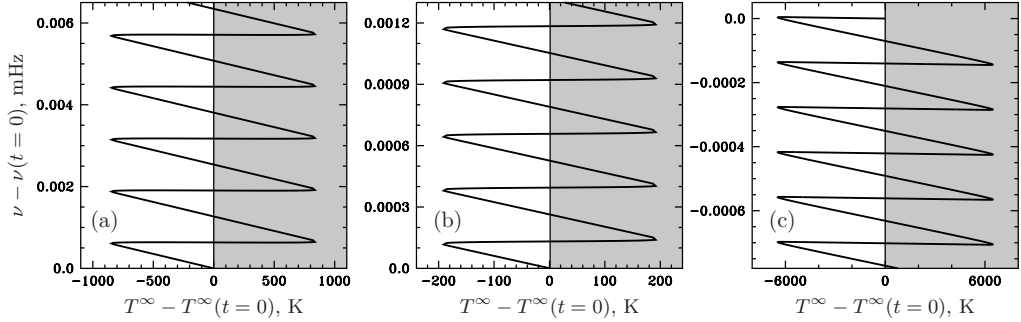}
    \end{center}
    \caption{Oscillations of the evolutionary track near the instability curve.
        Shown are variations of the rotation frequency (in millihertzs) and internal temperature (in Kelvins)
        starting from some initial moment of time $t=0$,
        for the following sources:
    (a) 4U 1608-522, the time interval is $\sim 3\,\rm years$,
		without Durca;
    (b) IGR J00291-5934,  the time interval is $\sim 45\,\rm years$,
		without Durca;
    (c) again IGR J00291-5934, the time interval is $\sim 2\,\rm years$,
	now with $\Lambda$-hyperonic Durca allowed for $r \leq 0.3R$,
   see 
	the text
	for details.}
    \label{Fig:evolution2}
\end{figure}

In Fig.\ \ref{Fig:spin} we show examples of
time
dependencies of the r-mode amplitude $\alpha$ (top panels), spin frequency $\nu$ (middle panels) and its derivative $\dot{\nu}$ (bottom panels).
Three left panels show these functions for the pulsar 4U 1608-522,
assuming no Durca processes operating in the star;
middle panels --- for the pulsar IGR J00291-5934 again without Durcas;
right panels --- for the same pulsar IGR J00291-5934, but with $\Lambda$-hyperonic Durca
allowed in the central region of a star with the radius $R_{\rm D}=0.3R$.
One can see that the pulsar 4U 1608-522 exhibits short periods (several days)
of rapid spin down due to the action of the $r$-mode torque,
which are repeated every $\widehat{P}=0.63\,\rm yr$.  
\footnote{The accurate treatment of the accretion episodes can
violate
the strict periodicity, but will not affect other results.
The same is true for other numerical examples considered here.}
When Durca processes are forbidden,
every $\widehat{P}=9.0\,\rm yr$ the pulsar IGR J00291-5934   
exhibits periods (lasting several dozens of days)
of enhanced $r$-mode torque, which results
in some decrease of NS spin up rate
(the torque is too weak to spin down the star).
Enhanced cooling makes the ``anti-glitches'' more pronounced
and shortens the time $\tau$ of violent spin down as well as
$\alpha$-oscillation period,
because $\widehat{P}\approx (33\div 48) \tau \propto 1/F_2(T_0)^{1/2}$ [see Equations (\ref{PPP}) and (\ref{tau})].
Thus, when $\Lambda$-hyperonic Durca processes
are open in the central region of a star,
the pulsar IGR J00291-5934 exhibits several days spin-down episodes
with well pronounced ``anti-glitch''.
$\alpha$-oscillation period in this case equals $\widehat{P}=0.35\,\rm yr$.  
One can see that $\alpha$-oscillation periods obtained
in numerical calculations (and given above) are slightly lower
than those presented in the Tables and
calculated with the approximate formula (\ref{PPP}).
The corresponding evolutionary tracks of the three
particular NS models described here are presented in three panels of Fig. \ref{Fig:evolution2}.

One more thing should be mentioned in relation with the Tables described above.
From Table \ref{Tab:param3} it follows that the high-temperature sources,
which are assumed to climb one of the two high-temperature peaks
(centered at $10^8\,\rm K$ and $1.5\times 10^8\,\rm K$) require a really high r-mode amplitude
to maintain their temperature {\it if} Durca processes
are allowed in the substantial part of their cores.
However, high values of r-mode amplitude result in the strong NS deceleration
(see the sixth column in Table \ref{Tab:param3})
that could be ruled out by timing analysis (see \citealt{ms13}).
Moreover, such a strong neutrino cooling (and hence NS deceleration) rises a question:
How did an NS get to the observed frequency?
In our scenario a star typically gets there by climbing up a stability peak
%
\footnote{This condition is not necessary
if we could allow for a possibility
that an NS has never climbed up a peak but only descended it.
This could be a reasonable assumption
for not too rapidly rotating NSs ($\nu \la 400\,\rm Hz$).}.
%
Assuming that an NS has climbed the peak up to the observed value of $\Omega$
we automatically
imply
that the r mode braking torque was not too strong,
so that a sufficiently high accretion rate
(e.g., $\dot{M}_{\rm acc\,high}\sim 10^{-8} M_\odot/\rm yr$ with corresponding $\dot{\Omega}_{\rm acc\,high}$),
that probably took place in the past, could spin the star up.
This assumption imposes the following
``spin up'' upper limit on  $\alpha_0$ at the stage of strong accretion,
when NS was spinning up,
\begin{equation}
\alpha_{\rm su} =\sqrt{\frac{\dot{\Omega}_{\rm acc\,high}}{G(T_0,\Omega_0) \Omega_0}},
\end{equation}
where we used Equation (\ref{Omega0temp}).
The value of $\alpha_{\rm su}$ is about $4\times10^{-6}$ for $\nu=600\,\rm Hz$
and scales with rotation frequency as
$\propto \nu^{-3.5}$
[this scaling follows from the Equations (\ref{tauGR2}), (\ref{Gdef}), (\ref{nuacc})
and the fact that $\tau_{\rm Diss}=|\tau_{\rm GR}|$ at the edge of the peak, see \citealt{gck14_large})
%
\footnote{Here, for definiteness, we employed the simplest expression for the accretion torque (\ref{nuacc}) that can be strongly modified by accounting for the magnetic field.}.
%
At the same time the thermal balance condition (\ref{Tequil}) dictates the value of $\alpha_0$.
For example, if 4U 1608-522 was accreting at $\dot{M}_{\rm acc\,high}\sim 10^{-8} M_\odot/\rm yr$
in the past, $\alpha_0$
for this source was $\alpha_0=1.30\times 10^{-5}$ if $R_{\rm D}=0.3R$.
%
\footnote{One can see that at the stage of strong accretion $\alpha_0$ was slightly lower than that calculated in Table \ref{Tab:param3} for the current accretion rate, the same is predicted by Equations (\ref{F2}) and (\ref{Tequil}).
However, the difference is very small,
because neutrino cooling with $R_{\rm D}=0.3R$ is much stronger
than the deep crustal heating rate even for $\dot{M}_{\rm acc\,high}\sim 10^{-8} M_\odot/\rm yr$
(the function $F_2(T_0)$ defined by (\ref{F2}) and entering the thermal balance condition (\ref{Tequil}),
is practically
independent of the accretion rate).
}
%
Thus $\alpha_0$ was higher than $\alpha_{\rm su}$, which means that 4U 1608-522 could not climb up the peak
at the current cooling rate.
There is, however, a possibility that Durca processes were closed in the past for the pulsar 4U 1608-522
and had opened (due to the mass accretion) only after this source
has climbed up the peak to a rather high frequency.
Only in the case of
such a fine tuning one can allow the Durca processes to operate
in a substantial part of 4U 1608-522 currently.
It is clear from the consideration above that
detailed analysis of the hottest NSs in LMXBs
could put tight constraints on
the neutrino emission due to Durca processes in the cores of these stars.

\subsection{Gravitational radiation}
\label{subSec:gravrad}

First, let us discuss whether gravitational waves
can, in principle, be detected on long time scales
(involving large number of $\alpha$-oscillation periods, $t\gg\widehat{P}$).
Detection statistics respond
to integrated signal power (\citealt{owen10}),
thus to rms (root-mean-square) value of $\alpha$,
i.e. to the quantity $\alpha_0$ (see Section \ref{Sec:Equations}).
Since $\alpha_0$ is defined by approximate thermal balance of a star, Equation (\ref{Tequil}),
the analysis in this case should be essentially the same as that of \cite{ms13},
where it was assumed that an NS undergoes r-mode oscillations at some (small) saturation amplitude
and is in thermal equilibrium during these oscillations.

Assume that we know the frequency
of gravitational waves $\omega$,
which equals to the r-mode frequency in the inertial frame,
and is given, in the Newtonian limit, by the formula,
$\omega=4/3\, \Omega$
(for $l=m=2$ r mode)%
\footnote{General relativistic corrections modify
this relation and introduce substantial uncertainties in $\omega$ (see, e.g., \citealt{ajh14,ioj15} and references therein).
However, direct measurements of NS spin frequency $\Omega$
and frequency $\omega_{\rm X}$
of X-ray emission modulation from the NS surface
due to perturbations of the surface by r mode
would allow to determine $\omega$ as $\omega=\omega_{\rm X}-2\Omega$.
We probably already have one such candidate
with both frequencies measured,
this is the pulsar XTE J1751-305, see \cite*{sm14} (see also \citealt*{sm14a}).}.
In this case gravitational waves can be detected if
(see, e.g., equation (6) of \citealt*{Watts_et_al_08} and \citealt*{jks98})
\begin{equation}
\left(\frac{1}{t}\int_0^t h_0(t')^2{\rm d}t'\right)^{1/2}\sqrt{t}=\left(\int_0^t h_0(t')^2{\rm d}t'\right)^{1/2}>11.4 \sqrt{S_{\rm n}(\omega)},
\label{hhh}
\end{equation}
where $S_{\rm n}(\omega)$ is the power spectral density
of the detector noise;
$t$ is the duration of signal collection;
$h_0(t')$ is the time-dependent intrinsic strain amplitude, which equals (see, e.g., \citealt{owen10})
\begin{equation}
h_0(t')=\sqrt{\frac{8 \pi}{5}}\,\frac{1}{d}\omega^3 \alpha(t'-\Delta t) M R^3 \widetilde{J},
\label{hhh2}
\end{equation}
where $d$ is the distance to the source, and
$\Delta t$ is the duration of gravitational signal propagation from the source to the detector; $c=G=1$ in this formula ($G$ is the gravitational constant).
For long-term observations (involving large number of oscillation periods of $\alpha$, $t\gg\widehat{P}$)
the rms value of $\alpha$ equals $\alpha_0$,
and $h_0(t')$ in formula (\ref{hhh})
can be replaced with its rms value $\left\langle h_0\right\rangle$,
given by equation (\ref{hhh2})
with $\alpha(t'-\Delta t)=\alpha_0$.
Obviously, the higher $\alpha_0$ is the more favorable is detection.
$\alpha_0$ can be rather high
(and detectable by, e.g., Advanced LIGO)
for most of the sources, especially if Durca processes are operating in their cores, see Tables \ref{Tab:param1}--\ref{Tab:param3}.
However, too high values of $\alpha_0$
would result in a very rapid spin-down of an NS,
and may contradict the timing measurements (future or existing) of NSs in LMXBs.
See detailed analysis of this issue in \cite{ms13}.

The registration of gravitational waves
during a shorter period of time,
particularly when $\alpha(t'-\Delta t)$, and hence $h_0(t')$ stays near its maximum
in the course of oscillations,
cannot improve statistics,
since any collection of even weak signal
contributes to
the statistics [see Equation (\ref{hhh})].
However, registration of the signal near the maximums of $\alpha$
could substantially decrease the time cost (or computational cost)
of the observation with the minimal losses in sensitivity.
Moreover, under certain conditions, the period of $\alpha$-oscillations,
$\widehat{P}$, can be pretty large, up to $10$ years
(see the third columns in Tables \ref{Tab:param1}--\ref{Tab:param3}).
In this case the duration of observations will be inevitably shorter than $\widehat{P}$.
Then, if the period of observations is chosen properly
(its duration is of the order of $\tau$ and $\alpha$ reaches its maximum during the period of observation),
the gravitational wave can be detected if [see Equation (\ref{hhh})]
\begin{equation}
\left(\int_0^t h_0(t')^2{\rm d}t'\right)^{1/2}\sim h_{0\,\rm max} \sqrt{\tau}\sim \left\langle h_0\right\rangle\sqrt{\widehat{P}}>11.4 \sqrt{S_{\rm n}(\omega)},
\label{hhh3}
\end{equation}
where $h_{0\,\rm max}$ is given by (\ref{hhh2}) with
$\alpha(t'-\Delta t)=\alpha_{\rm max}$. To derive
(\ref{hhh3}) we used the fact that most of the
$\alpha$-oscillation period $\alpha$ is negligibly small
and r mode excites for a time of the order of $\tau$, and
adopted an approximate equality (\ref{rough}). Notice that
$\widehat{P}$ in Equation (\ref{hhh3}) can be as large as
several years, but one does not need to process the signal
collected during the whole period $\widehat{P}$, it is only
necessary to process the signal collected during the time
interval $\tau$, which is $\sqrt{2(k^2-\log k^2)}\,k\approx
(33\div 48)$ times shorter than $\widehat{P}$. The values
of $\left\langle h_0\right\rangle\sqrt{\widehat{P}}$ are
presented in the last column of Tables
\ref{Tab:param1}--\ref{Tab:param3}.

To choose the correct time for the detection of gravitational signal
we either should be sufficiently lucky or
need 
to know $\alpha$-oscillation parameters
(when exactly r-mode amplitude reaches the maximum).
These parameters can, in principle, be found from the
peculiarities in timing behavior of a given source,
see Section \ref{subSec:timing}.

\section{$\alpha$-oscillations vs existing observations}
\label{Sec:evidences}

Since
gravitational waves have not yet been detected, the only observational
manifestation of $\alpha$-oscillations in existing
observations can be found in timing of NSs.

Some of the NSs in LMXBs are X-ray pulsars showing
pulsations
from time to time
(during outbursts). For this reason
their timing parameters are poorly constrained. The best
measurements of timing parameters were done for the source
HETE J1900.12455 (see \citealt{Patruno12}). This source showed a
strong decrease
in
the rotation frequency time derivative,
$\dot{\nu}$, on a short timescale of the order of $26\pm 4$
days. \cite{Patruno12} fitted this change of
$\dot{\nu}$ as an exponential decay and interpreted it as a
manifestation of the magnetic field burial. However, the
same observational data can also be interpreted (and
fitted) as $\alpha$-oscillations discussed in this paper.
Unfortunately, the quality of the observational data does
not allow
one
to distinguish which model is correct and
which is not (private communication with A.~Patruno).
Additional X-ray observation of this source with high
temporal resolution would be highly desirable.

On the other hand,
the timing
behavior of millisecond radio pulsars is often measured very
precisely and an upcoming projects like SKA (\citealt*{ks15}; \citealt{tauris15}) will increase the number of such pulsars.
The scenario proposed by
\cite{gck14_short,gck14_large} admits that after accretion
stage finishes and an NS becomes a millisecond radio pulsar
it may still stay attached to one of the stability peaks
(there are, however, other possibilities, see
\citealt{gck14_short,gck14_large}). This peak should be a
low-temperature one, because observations indicate that
non-accreting millisecond pulsars are
relatively cool.%
\footnote{
Due to this reason their surface temperatures are still
not measured (only X-ray emission from
polar caps is registered for some of them).
The only exception is the
pulsar PSR~J0437$-$4715 (spin frequency $\nu\approx
173.7$~Hz) with the redshifted surface temperature
$(1.25\div 3.5)\times 10^5\,\rm K$,
corresponding to
the emitting radius $15\div 7.8$~km
(\citealt{durant12}); even the largest estimate of the surface
temperature for this pulsar corresponds to $T^\infty<2\times 10^7$~K, which
is too low to make PSR~J0437$-$4715 r-mode unstable.}
%
NS, attached to the peak, may experience
$\alpha$-oscillations. However, to our best knowledge,
there are no known peculiarities in timing behavior of
millisecond pulsars, which can be attributed to the timing
features discussed here.
We can not exclude that these peculiarities
are masqueraded by the pulsar timing noise.
The large periods of $\alpha$-oscillations and small amplitudes of
the accompanying ``anti-glitches'' in cold NSs with low
$L_\mathrm{cool}$ argue in favour of this assumption%
\footnote{Note that the most stable millisecond pulsars
(whose timing noise is extensively studied
in relation to the problem of gravitational wave detection)
are not very rapid rotators
and
can be stable with respect to r-mode excitation
(for example, the fastest
pulsar in NANOGrav project, PSR~J1909--3744,
has $\nu\approx 339$~Hz, see, e.g., \citealt{demorest_etal13}).}.
However, if timing of all millisecond pulsars is indeed
unaffected by $\alpha$-oscillations,
this can indicate
that either ($i$) all millisecond pulsars are stable with
respect to r-modes (for example,
due to evolution in a nontrivial r-mode
instability window, see, e.g., appendix D in
\citealt{gck14_large} or due to some unknown dissipation mechanism, which
makes the
r-mode instability irrelevant for the physics of NSs) or ($ii$) all
unstable millisecond pulsars are pinned to the stability
peaks, which are sufficiently wide, so that
$\alpha$-oscillations do not excite for them
(see a corresponding discussion after the expression
\ref{stability} in Section \ref{Sec:Equations})
\footnote{Formally, there
is an additional possibility that the unstable millisecond
pulsars are not associated with the stability peaks.
However, it requires extremely small r-mode saturation
amplitudes (see, e.g., \citealt{as15}), which seems to be very
unlikely, see a footnote 9 in \cite{gck14_large}.}.
The latter assumption looks very natural,
because the low-temperature peaks should be wider than the
high-temperature ones, as is confirmed by calculations of
the spectra of non-rotating superfluid NSs
(\citealt{gkcg13,gkgc14}). It is interesting to ask how wide
should be a peak to allow an NS to climb it up/down without
oscillations?
For example, to make the motion of IGR J00291-5934
along the left edge of the stability peak stable
(i.e., to make $\gamma(A_0)<0$)
it is enough to increase the coupling parameter $s$ to the value $s=0.04$
(by a factor of $4$),
which looks to be realistic
(this estimate is made for the case when Durca is open in the region
$r \leq R_{\rm D}=0.1R$).

Even if non-accreting millisecond
pulsars are unaffected by $\alpha$-oscillations, some NSs
in LMXBs (accreting millisecond pulsars), which are
attached to sharper peaks centered at higher temperatures,
still can be unstable with respect
to $\alpha$-oscillations and thus should exhibit them.
To check this possibility we need more X-ray observations with high temporal
resolution that can be achieved by future missions such as
LOFT (\citealt{LOFT}), NICER (\citealt{NICER}), and SRG
(\citealt{SRG}).
Another opportunity to detect $\alpha$-oscillations can
be associated with the recently discovered transitional
millisecond pulsars,
which are switching between rotation-
and accretion-powered states
[currently, three of such objects are known:
PSR\ J1023$+$0038 ($\nu\approx 592.4$~Hz,
\citealt{Archibald_etal09,Stappers_etal14}), IGR\
J18245$-$2452 ($\nu\approx 254.3$~Hz,
\citealt{Papitto_etal13}), and XSS\ J12270$-$4859
($\nu\approx 593.0$~Hz, \citealt{Roy_etal15})].
In a rotation-powered state these NSs are observed as
radio pulsars,
giving a chance to study their timing properties precisely.
The accretion-powered state guaranties that these
pulsars can not be too cold, and thus likely to be affected
by r-mode instability and $\alpha$-oscillations. For
example, the red-shifted surface temperature of PSR J1023$+$0038
can be as large as $5\times10^5$~K
(\citealt{Homer_etal06,Bogdanov_etal11}). It corresponds to
the internal temperature $T^\infty=(1.5\div3)\times10^7$~K,
implying that PSR J1023$+$0038 can be on the stability
peak. Its timing behavior was reported as `complex' by
\cite{Archibald_etal13} and can be influenced by $\alpha$-oscillations.
The reliable registration of $\alpha$-oscillations would
confirm the crucial role of the resonance interaction of
oscillation modes in the evolution of NSs in LMXBs
(\citealt{gck14_short,gck14_large}). The measurements of
$\alpha$-oscillation parameters (such as the period of
oscillations, $\widehat{P}$) would provide us with a new
powerful tool to constrain the properties of superdense
matter.

\section{Conclusions}
\label{Sec:Conclusions}
We showed that when an NS climbs the stability peak
according to the
scenario proposed by \cite{gck14_short, gck14_large},
its parameters can undergo {\it nonlinear} oscillations near their equilibrium values
($\alpha$-oscillations).
We studied these oscillations analytically and
determined their key parameters such as the oscillation period and amplitude.
We also found a
rigorous criterion showing when
$\alpha$-oscillations
become unstable.
For fully developed unstable $\alpha$-oscillations we demonstrated that,
most of the $\alpha$-oscillation period,
the amplitude of r-mode $\alpha$
stays negligibly small
and increases only for a short time of the order of hours-months (depending on parameters)
by a factor of about $5\div 6$ in comparison to its root mean square value.
Thus, the spin frequency derivative,
which is not affected by r modes most of the time,
dramatically decreases, generally approaching large negative values,
when $\alpha$ becomes large.
This results in small ``anti-glitches''
(for NSs in LMXBs considered here
their sizes lie within the range
$5.3\times 10^{-12}<|\Delta \nu/\nu| < 2.3\times 10^{-8}$
for the adopted parameters),
which typically last hours-months depending on the model.
This fact can substantially affect timing analysis
of rapidly rotating NSs (radio-, X-ray and transitional millisecond pulsars),
in particular it potentially can be useful for interpretation of timing behavior of NSs in LMXBs,
such as HETE J1900.12455 (\citealt{Patruno12}) and IGR J00291-5934 (\citealt{Patruno10}).
Moreover,
the very existence of
oscillations of the r-mode amplitude $\alpha$
could substantially decrease the time cost of gravitational signal detection from LMXBs,
because collection of the signal during the time period $\tau$
when $\alpha$ is close to its
maximum (typically, hours-months)
results in almost the same sensitivity
as collection of the signal
during the whole $\alpha$-oscillation period ($\sim 40\tau$).
The identification of $\alpha$-oscillations in observations would confirm the scenario of NS evolution in LMXB
proposed in \cite{gck14_large,gck14_short} and would show that
r modes are
crucially important for understanding of
rapidly rotating NSs.
It would also open a new independent way to constrain the properties of superdense matter of NS cores by measuring $\alpha$-oscillation parameters.

\appendix
\section{Cooling rate of an NS}
\label{Sec:Appendix}

To calculate the luminosity
in the absence of Durca processes (let us denote it $L_{\rm noDurca}$)
we use an approximate formula, which expresses $L_{\rm noDurca}$
as a function of the internal (redshifted)
stellar temperature $T^\infty$, see formula (A1) in appendix A of \cite{gck14_large}.
This formula fits numerical results for the luminosity obtained with
the relativistic cooling code,
described in detail by \cite{gkyg04,gkyg05} and \cite{yp04}.
To calculate $L_{\rm noDurca}$,
essentially the same microphysics input was used
as in \cite{gkyg04},
namely, the parameterization by \cite{hh99}
of APR EOS (\citealt{apr98}) was employed,
and an NS with the mass $M=1.4 M_{\odot}$ was considered.

It is generally believed that
hyperons appear in the NS matter at densities $\sim (2-3) \rho_0$,
where $\rho_0=2.8\times 10^{14}$~g~cm$^{-3}$ is the nuclear density
(see, e.g., \citealt{wcs12,bhzbm12,ghk14} and references therein).
This threshold for hyperon appearance is rather low
which means
that substantial fraction of NSs (probably) host hyperons in their cores.
Most of the modern models predict that the first hyperons
to appear with increasing density is $\Lambda$ hyperons (see, e.g., \citealt{wcs12,bhzbm12,ghk14}).
It is quite likely that the superfluid gap for them
is very low (\citealt{tmc03,tnyt06,ws10}) so that
$\Lambda$ hyperons are normal
at temperatures relevant to NSs in LMXBs.
According to many microscopic models, protons are also non-superconducting
at sufficiently high densities (e.g., \citealt{plps09}).
As soon as $\Lambda$ hyperons
appear, the
very powerful direct Urca process,
$\Lambda \rightarrow p+ l + \widetilde{\nu_l}$, and the inverse one,
$p+ l  \rightarrow \Lambda+ \nu_l$,
start to operate in the NS interiors
(here $l$ stands for a lepton, electron or muon).
The corresponding neutrino emissivity was calculated by \cite{pplp92,ykgh01}.
Here we present the total emissivity due to the reactions with electrons and muons
and assume that it is not suppressed by the proton
and/or $\Lambda$ hyperon superfluidity.
\begin{equation}
Q\approx 8\times 10^{27} \left(\frac{n_e}{n_0}\right)^{1/3}\frac{m_\Lambda^\star m_{\rm p}^\star}{m_{\rm n}^2} T_9^6 r_{\rm \Lambda p}\,\rm erg\, cm^{-3}\, sec^{-1},
\label{QQ}
\end{equation}
where
$n_0=0.16$~fm$^{-3}$ is the nuclear number density,
$n_e$ is the electron number density,
$T_9\equiv T/10^9\,\rm K$,
$r_{\rm \Lambda p}=0.039$,
$m_\Lambda^\star$,
and $m_{\rm p}^\star$ are the effective masses
for $\Lambda$ hyperons and protons, respectively,
$m_{\rm n}$ is the mass of a free neutron.
In all numerical calculations we adopt
$m_\Lambda^\star=0.7m_\Lambda$ ($m_\Lambda=1115.63\, \rm MeV$)
and $m_{\rm p}^\star= 0.7 m_{\rm n}$.
As long as the emissivity is not too sensitive
to the actual value of the electron number density $n_{\rm e}$,
we assume $n_{\rm e}=0.25 n_0$
throughout the whole region where the Durca processes are open
(it is a good approximation for the central part of a star).
This is a typical value of $n_{\rm e}$
near the threshold density of $\Lambda$ hyperon appearance (\citealt{ghk14}).

Using $Q$ from the equation (\ref{QQ})
we calculate (or better to say, estimate)
the neutrino luminosity due to the $\Lambda$ hyperon Durca processes as
\begin{equation}
L_{\rm Durca}=Q \frac{4}{3}\pi R_{\rm D}^3,
\end{equation}
where $R_{\rm D}$ is the radius of the inner part of the stellar core
where $\Lambda$ hyperons are present;
we treat $R_{\rm D}$ as a free parameter and consider three cases:
$R_{\rm D}=0$ (no $\Lambda$ hyperons),
$R_{\rm D}=0.1 R$, and $R_{\rm D}=0.3 R$.
The resulting luminosity (cooling rate) $L_{\rm cool}$ is calculated
as a sum of $L_{\rm noDurca}$ and $L_{\rm Durca}$, $L_{\rm cool}=L_{\rm noDurca}+L_{\rm Durca}$.

\section*{Acknowledgments}

We are grateful to Alessandro Patruno for correspondence
and some clarifying comments on the timing behavior of HETE J1900.12455
and to Crist$\rm{\acute{o}}$bal Espinoza and Y.A. Shibanov for discussions.
This study was partially supported
by RFBR (grants 14-02-00868-a and 14-02-31616-mol-a),
and by RF president programme
(grants MK-506.2014.2 and NSh-294.2014.2).


\label{lastpage}

\end{document}